\documentclass[fleqn,usenatbib]{mnras}

\usepackage{newtxtext,newtxmath}

\usepackage[T1]{fontenc}
\usepackage{ae,aecompl}
\usepackage{color, colortbl}

\usepackage{float}
\usepackage[caption = false]{subfig}
\usepackage{graphicx}	
\usepackage{amsmath}	
\usepackage{booktabs}
\usepackage{color}

\title[Voids in MG with massive neutrinos]{Cosmic voids in modified gravity models with massive neutrinos}

\author[S. Contarini et al.] {Sofia Contarini$^{1,2,3}$\thanks{E-mail:
    sofia.contarini3@unibo.it}, Federico Marulli$^{1,2,3}$, Lauro Moscardini$^{1,2,3}$, Alfonso Veropalumbo$^{4}$, \newauthor Carlo Giocoli$^{2,1,3}$ and Marco Baldi$^{1,2,3}$ \\~\\
    $^1$ Dipartimento di Fisica e Astronomia ``Augusto Righi'' -, Alma Mater Studiorum Universit\`{a} di Bologna, via Piero Gobetti 93/2, I-40129 Bologna, Italy\\ $^2$ INAF - Osservatorio di Astrofisica e Scienza dello Spazio di Bologna, via Piero Gobetti 93/3, I-40129 Bologna, Italy \\ $^3$ INFN - Sezione di Bologna, viale Berti Pichat 6/2, I-40127 Bologna, Italy \\ $^4$ Dipartimento di Fisica, Universit\`a degli Studi Roma Tre, via della Vasca Navale 84, I-00146 Roma, Italy}

\date{Accepted XXX. Received YYY; in original form ZZZ}

\pubyear{2020}

\begin{document}
\label{firstpage}
\pagerange{\pageref{firstpage}--\pageref{lastpage}}
\maketitle

\begin{abstract}
Cosmic voids are progressively emerging as a new viable cosmological
probe. Their abundance and density profiles are sensitive to
modifications of gravity, as well as to dark energy and neutrinos. The
main goal of this work is to investigate the possibility of exploiting
cosmic void statistics to disentangle the degeneracies resulting from
a proper combination of $f(R)$ modified gravity and neutrino mass. We
use N-body simulations to analyse the density profiles and size
function of voids traced by both dark matter particles and haloes. We
find clear evidence of the enhancement of gravity in $f(R)$
cosmologies in the void density profiles at $z=1$. However, these
effects can be almost completely overridden by the presence of massive
neutrinos because of their thermal free-streaming.
Despite the limited volume of the analysed simulations
does not allow us to achieve a statistically relevant abundance of
voids larger than $40 \ \mathrm{Mpc}/h$, we find that the void size function
at high redshifts and for large voids is potentially an effective probe to
disentangle these degenerate cosmological models, which is key in the
prospective of the upcoming wide field redshift surveys.

\end{abstract}




\section{INTRODUCTION}

The Universe has recently entered a phase of accelerated
expansion. This revolutionary discovery goes back to more than two
decades ago and was originally achieved thanks to distant type Ia
supernovae \citep{Riess1998, Perlmutter1999, Schmidt1998}. The
following observations of the cosmic microwave background anisotropies
and large-scale structures have then supported this scenario
\citep[e.g.][]{SDSS2005, Komatsu2011, Bennett2013, Planck2018}, ,
which is now widely accepted among the scientific
community. Nevertheless, the understanding of the physics behind this
accelerated expansion remains one of the fundamental open questions in
cosmology, and a plethora of theoretical models have been proposed to
explain this phenomenon \citep[see e.g.][for a review]{Yoo2012}.

The standard paradigm of modern cosmology, the $\Lambda$-cold dark
matter ($\Lambda$CDM) model, interprets the accelerating expansion of
the Universe as due to the existence of an extra component, named
\textit{dark energy}, contributing to about the $70\%$ of the total
energy density of the Universe. In its most trivial description, this
component behaves like a fluid with a negative equation of state,
which can be straightforwardly described by the cosmological constant
$\Lambda$ in the Einstein field equations. Thanks to its simplicity
and its concordance with the majority of current cosmological
observations, the $\Lambda$CDM model is nowadays the most popular and
widespread cosmological model \citep{Shafieloo2010,
Heavens2017}. However, this scenario has been often questioned,
since it clashes with both some theoretical and observational
issues. The former concern for instance the coincidence and the
fine-tuning problems \citep[][but see \citealt{Bianchi2010} for an
alternative perspective]{Weinberg1989, Sean2001, Martin2012}, while
the latter are raised in particular by the recent measurements of the
Hubble constant, $H_0$, together with other well-known anomalies and
tensions \citep[see e.g.][and references therein]{Bernal2016,
Moresco2017, Verde2019}. Hence, new ideas and different theoretical
approaches arose to solve or alleviate these possible fundamental
inconsistencies. Among the proposed solutions, some models interpret
the dark energy component as a dynamical variable slowly varying with
the cosmic time, or as exotic new forms of energy that would cause the
observed late time accelerated rate of the Universe \citep[see
e.g.][and references therein]{Frieman2008, Wen2018}. There are also
alternative explanations which involve a modification of General
Relativity (GR) in a manner that leads to accelerating solutions. In
this class of models the standard GR is supposed to be inadequate on
certain cosmological scales, which implies also the introduction of
new physical degrees of freedom in the gravitational theory to explain
its behaviour on small scales \citep[see e.g.][]{Dolgov2003,
Nojiri2006, Clifton2012, Joyce2015, Ishak2019}.  In particular,
Modified Gravity (MG) models tend to closely mimic the effect of the
cosmological constant on the expansion history of the Universe.  To
satisfy the solar system tests and the local high-precision
measurements \citep{LeVerrier1859, Bertotti2003, Will2005}, these
models have to introduce a \textit{screening mechanism}, that
basically recovers the predictions of standard GR on small scales
\citep{KhouryWeltman2004, Hinterbichler2010, Brax2013, Brax2014}.
Most viable MG models are quite degenerate at the background level and
can produce discernible features only through their effects on
structure formation at linear and nonlinear scales.
Additionally, it has been recently highlighted the
presence of strong observational degeneracies between the effects of
some of these models and those including massive neutrinos
\citep{He2013, Motohashi2013, Baldi2014, Wright2017,
Giocoli2018}. Neutrinos are indeed another elusive component of the
$\Lambda$CDM cosmology, and although the Standard Model of particle
physics assumes they are massless, the evidence of solar neutrino
oscillations proved they in fact possess a mass
\citep{Becker-Szendy1992, Fukuda1998, Ahmed2004}.

In this paper we will investigate the degeneracies emerging from a
proper combination of the parameters of the so-called $f(R)$ class of
MG models and of the total neutrino mass, $\sum m_\nu$. In particular,
\cite{Baldi2014} have demonstrated that many standard cosmological
statistics, as the nonlinear matter power spectrum, the halo abundance
and the halo bias, show a limited discriminating power for some
specific combinations of $f(R)$ gravity parameters and neutrino mass
values, for which they revealed to be statistically consistent with
the $\Lambda$CDM predictions. Our goal is to investigate whether it is
possible to disentangle these degenerate cosmological scenarios by
exploiting a novel promising cosmological probe, i.e. the cosmic
voids, and more precisely their radial profiles and abundances.

Voids are defined as large regions of the Universe with low-density
interiors and shallow gravitational potentials. Thanks to these unique
features, they constitute excellent laboratories for investigating the
implications of MG theories and the presence of massive
neutrinos. Indeed, screening mechanisms operate weakly within cosmic
voids, making them potentially more affected by the possible
deviations from GR \citep{Spoylar2013, Barriera2015, Voivodic2017,
Baker2018, Falck2018}. Furthermore, voids are
particularly sensitive to neutrinos \citep{ZengWhite1991,
Clampitt2013, Villaescusa-Navarro2013, Massara2015, Banerjee2016,
Hamaus2017, Schuster2019, Kreisch2019, Dvorkin2019, Wang2019,
Pisani2019}, since their typical sizes span the range of neutrino
free-streaming scales and the density fraction of neutrinos is more
prominent in these zones compared to those in high-density ones. Both
the void density profiles and void abundances have been shown to
possess a great potential in constraining cosmological parameters
\citep[see e.g.][]{Pisani2015, Hamaus2016, Hamaus2020, Sahlen2016,
Sahlen2018, Sahlen2019, contarini2019, Aubert2020}. Void profiles
show a characteristic shape that depends on the mean radius of the
void sample, and are well reproduced by the functional form provided
by \cite{Hamaus2014}. Void abundances have been studied by a number of
authors \citep[see e.g.][]{SVdW2004, Jennings2013, Pisani2015,
RonconiMarulli2017, Chan2019, Correa2020}, and have been recently
explored in relation to the modifications induced on the void size
function by the bias factor of the tracers \citep{Pollina2016,
Pollina2017, pollina2019, Roncarini2019, contarini2019, verza2019}.

In this work we explore the possibility of breaking the degeneracy
between MG and neutrino effects by analysing cosmic voids
identified by means of the void finder {\small VIDE}
\citep{SutterVIDE2015} in the DUSTGRAIN-\textit{pathfinder}
simulations \citep{Giocoli2018, Hagstotz2019}, a set of N-body
simulations with a volume of $(750 \ \mathrm{Mpc}/h)^3$, including cosmological models with both
$f(R)$ gravity and massive neutrinos.
We study voids traced by both the distribution of DM
particles and of collapsed haloes, measuring their density profiles
and their abundance as a function of void sizes. The latter is then
compared to the theoretical model provided by \cite{Jennings2013}, and
modified to include the effect of the tracer bias
\citep[see][]{Roncarini2019, contarini2019}. For a proper comparison
with the theoretical predictions, the void sample is analysed according to
the prescriptions introduced in \cite{RonconiMarulli2017}.

The paper is structured as follows. In Section \ref{sec:all_theories}
we introduce the theoretical context in which this analysis is
inserted, summarising some of the fundamental notions needed to
understand the effects of $f(R)$ MG and massive neutrinos.
Then we present the theoretical model for the
void size function, together with the prescriptions required to take
into account the modifications induced by the usage of biased tracers
to identify the voids. In Section \ref{sec:catalogue_preparation} we
describe the set of simulations analysed in this work and the methods adopted to build the catalogues of haloes and voids.
We also outline the cleaning procedure applied to recover
cosmic voids consistently with the assumptions used in their theoretical modelling. In Section
\ref{sec:results} we present the results of our
analysis, discussing the void density profiles and void abundances,
and focusing on their possible exploitation to disentangle the cosmic degeneracies
previously introduced. In Section \ref{sec:conclusions} we finally
summarise the main conclusions drawn in this paper.


\section{THEORETICAL MODELS} \label{sec:all_theories}

This Section presents the fundamental theoretical
background on which this work is based. We briefly start by introducing
the MG and massive neutrino models characterising the cosmological simulations
analysed in this work. Then we present and discuss the theoretical
model for the void size function.

\subsection {\texorpdfstring{$f(R)$}{TEXT} modified gravity theory}
\label {sec:MG_theory}

Among the variety of cosmological models proposed to explain the accelerated expansion of the Universe, we consider those alternative theories that imply a deviation from the standard behaviour of gravity by modifying the left hand side of the Einstein's field equations\footnote{We adopt natural units, thus $c=\hbar=1$.}:
\begin{equation}
  \label{eq:einstein_eq}
  R_{\mu\nu}-\frac{1}{2}g_{\mu\nu}R = 8 \pi GT_{\mu\nu}\, ,
\end{equation}
in which $R_{\mu\nu}$ is the Ricci curvature tensor, $R$ the Ricci scalar, $g_{\mu\nu}$ the metric tensor, $G$ the Newton's
gravitational constant and $T_{\mu\nu}$ the stress-energy tensor. One
of the simplest ways to modify the GR equations is to change the
Einstein-Hilbert action, $S$, introducing a function $f$ of the Ricci
scalar:
\begin{equation}
  \label{eq:mg1}
    S = \int \mathrm{d}^4x \sqrt{-g} \  \Big(\frac{R+f(R)}{16 \pi G}\Big)
    + \mathcal{L}_m\, ,
\end{equation}
where $\mathcal{L}_m$ is the Lagrangian density of all matter
fields. In this class of MG theories, called $f(R)$ models, GR is recovered
by imposing $f$ to be proportional to the cosmological constant $f =
-2\Lambda^{\mathrm{GR}}$. An important requirement to be satisfied by
the function $f(R)$ is to match the observed $\Lambda$CDM expansion
of the Universe on large scales and to evade the solar system
constraints at the same time. A valid form of this function,
introduced by \citet{Hu&Sawicki2007}, is the following:
\begin{equation}
  \label{eq:mg2}
    f(R) = -m^2 \frac{c_1 \Big( \frac{R}{m^2} \Big)^n}{c_2 \Big( \frac{R}{m^2} \Big)^n+1}\, ,
\end{equation}
where $m^2 \equiv H_0^2 \Omega_\mathrm{M}$ defines the mass scale $m$, while $c_1$, $c_2$ and $n$ are non-negative free parameters of the model. In particular we refer to the case in which $c_1/c_2 = 6 \Omega_\Lambda/ \Omega_\mathrm{M}$, where $\Omega_\Lambda$ and $\Omega_\mathrm{M}$ represent the present vacuum density and matter density parameters, respectively. 
Under this specific condition the background expansion history is indeed consistent with the one predicted by the $\Lambda$CDM model. Moreover, imposing $c_2(R/m^2)^n \gg 1$ the scalar field $f_R \equiv d f(R)/dR$ can be approximated by:
\begin{equation}
  \label{eq:mg4}
    f_R \approx -n \frac{c_1}{c_2^2} \bigg( \frac{m^2}{R} \bigg)^{n+1}\, .
\end{equation}
In this work we restrict our analysis to the case $n=1$. With this
choice the scalar field can in fact be expressed by means of the
parameter $c_2$ only, and the model at the present epoch can be
represented by the parameter $f_{R0}$:
\begin{equation}
  \label{eq:mg5}
    f_{R0} \equiv -\frac{1}{c_2} \frac{6 \Omega_{\Lambda}}{\Omega_\mathrm{M}} \bigg( \frac{m^2}{R_0} \bigg)^2\, ,
\end{equation}
where $R_0$ indicates the background value of the Ricci scalar at the present time.
Now we can derive the modified Einstein equations by varying the action defined in Eq. \eqref{eq:mg1} with respect to the metric $g_{\mu\nu}$: 
\begin{equation}
  \label{eq:mg6}
    f_R R_{\mu\nu} - \frac{1}{2}fg_{\mu\nu}-\nabla_\mu \nabla_\nu f_R + g_{\mu\nu} \Box f_R = 8 \pi G T_{\mu\nu}\, ,
\end{equation}
in which $\nabla$ is the covariant derivative and $\Box$ is the
D'Alembert operator defined as $ \Box \equiv g^{\mu\nu} \nabla_\mu
\nabla_\nu$. Here $f_R$ turns out to be the responsible for the
modification of the GR theory and plays the role of a new dynamical
scalar degree of freedom. From the trace of Eq. \eqref{eq:mg6} we can
obtain the equation of motion for this scalar field:
\begin{equation}
  \label{eq:mg7}
    \nabla^2 \delta f_R = \frac{a^2}{3} \left[ \delta R (f_R) - 8 \pi G \delta \rho_m \right]\, ,
\end{equation}
where $a$ is the scale factor of the metric. To obtain the equivalent of the Poisson equation for the scalar metric perturbation ${2\psi = \delta g_{00}/g_{00}}$, we extract the time-time component from Eq. \eqref{eq:mg6}:
\begin{equation}
  \label{eq:mg8}
    \nabla^2 \psi = \frac{16 \pi G}{3} a^2 \rho_m -\frac{a^2}{6} \delta R(f_R)\, ,
\end{equation}
assuming small perturbations on a homogeneous background\footnote{$\delta f_R \equiv f_R - \bar{f_R}$, $\delta R \equiv R - \bar{R}$ and $\delta \rho_m \equiv \rho_m - \bar{\rho}_m$, where the barred values represent the background quantities.} and a slow variation for $f_R$ (quasi-static field). 

From Eqs. \eqref{eq:mg7} and \eqref{eq:mg8} it is possible to derive the exact solution for the extreme cases $|f_{R0}| \gg |\psi|$ and $|f_{R0}| \ll |\psi|$. It can be demonstrated that, when the field is large, thus in the former case, the Compton wavelength of the scalar field ${\mu^{-1} = (3 \ \mathrm{d}f_R/\mathrm{d}R)^{1/2}}$ determines the interaction range of an additional fifth force, which can enhance the gravity field up to a factor of $4/3$ for $k \gg \mu$. Standard gravity is instead restored for scales $k \ll \mu$. In the latter case, instead, the value of $f_{R0}$ is small and Eq. \eqref{eq:mg8} can be approximated by the standard Poisson equation, leading to the recovery of GR in regions of high space-time curvature. This is the so-called \textit{Chameleon screening mechanism}, which has the effect of hiding the additional fifth force on small scales, suppressing its strength inside large matter overdensities. By solving Eq. \eqref{eq:mg7} under the assumption of small perturbations in the homogeneous background, $\delta f_R \le \bar{f_R}$, we can obtain the screening condition for an ideal spherical source of mass $M$ causing the fluctuation of the scalar field:
\begin{equation}
  \label{eq:mg9}
    |f_R| \le \frac{2}{3}\psi_N(r)\, ,
\end{equation}
where $\psi_N=GM/r$ is the Newtonian potential of the overdensity. In this approximation, the enhancement of gravity is carried out only by the distribution of mass outside the radius for which ${\psi_N(r) = 3/2 \ |f_R|}$, that constitutes the transition point between the screened and un-screened regimes.

We can now assess valid estimations for the free parameter $f_{R0}$. The case in which $f_{R0} \ll \psi_N$ has no relevant cosmological interest since the fifth force is always screened, hence the resulting scenario is indistinguishable from GR even on large scales. On the other hand, for $f_{R0} \gg \psi_N$, we would face the implausible situation in which gravity is always enhanced. Therefore the parameter $f_{R0}$ should be settled around the same order of magnitude of the Newtonian potential $\psi_N$, that in turn typically shows values in the range ${10^{-5} \le \psi_N \le 10^{-6}}$.


\subsection {Degeneracies with massive neutrinos}

Neutrinos are massive particles participating to the total matter
content of the Universe and to the growth of cosmic
structures. Given their small masses,
neutrinos decouple from high relativistic particles at the early
stages of the Universe, when their thermal energy drops below their
mass. Precision cosmology allows nowadays to put strong
constraints on their physics and especially on the sum of their mass
eigenstates $m_\nu \equiv \sum_i m_{v_i}$.
The total neutrino mass is indeed constrained by several
astronomical observations to be ${m_\nu \lesssim 0.1 - 0.3
\ \mathrm{eV}}$ \citep[see e.g.][]{Seljak2006, Riemerorensen2013,
Lu2015, Lu2016, Cuesta2016, Kumar2016, Yeche2017, Poulin2018}, and
their contribution to the total amount of energy in the Universe at
late cosmological epochs can be computed as \citep{Mangano2005}:
\begin{equation}
  \label{eq:neutrino_mass}
    \Omega_\nu \approx \frac{m_\nu}{93.14 \ h^2 \mathrm{eV}}\, .
\end{equation}
Given their weak interaction cross-section, neutrinos can be considered as a DM component. However, contrary to CDM particles, neutrinos can free-stream from high density perturbations of matter thanks to their high thermal velocity. Indeed we can derive the typical scales travelled by neutrino perturbations, described by the free-streaming length:
\begin{equation}
  \label{eq:free-streaming_length}
    \lambda_\mathrm{FS}(z,m_v) = a(z) \frac{2\pi}{k_\mathrm{FS}} = 7.7(1+z) \frac{H_0}{H(z)} \bigg( \frac{1\mathrm{eV}}{m_v}\bigg) \ \mathrm{Mpc}/h \, ,
\end{equation}
where $k_\mathrm{FS}$ is the associated free-streaming wavenumber, which during the neutrino non-relativistic transition, $z_\mathrm{nr}$, reaches the minimum value \citep{Lesgourgues2013}:
\begin{equation}
  \label{eq:free-streaming_wavenumber}
    k_\mathrm{FS}(z_\mathrm{nr}) \simeq 0.0178 \ \bigg(\Omega_\mathrm{M} \frac{m_v}{\mathrm{eV}} \bigg) \ h/\mathrm{Mpc}  \, .
\end{equation}
Therefore modes with $k < k_\mathrm{FS}$ evolve as CDM perturbations
since neutrino velocities can be neglected, while on small scales ($k
\gg k_\mathrm{FS}$) free-streaming leads to the slowdown of the
neutrino perturbation growth.  Besides suppressing the clustering
below their thermal free-streaming scale, neutrinos also affect the
shape of the matter auto-power spectrum \citep{Brandbyge2008,
Saito2008, Saito2009, Brandbyge2009, Brandbyge2010b, Agarwal2011,
Wagner2012}, the halo mass function \citep{Brandbyge2010,
Marulli2011, Villaescusa-Navarro2013b}, the scale-dependent bias
\citep{Chiang2019}, the clustering properties of CDM haloes and
redshift-space distortions \citep{Viel2010, Marulli2011,
Villaescusa-Navarro2014, Castorina2014, Castorina2015, Zennaro2018,
Jorge2019}, and also the number counts and profiles
of cosmic voids (see Section \ref{sec:results}).

It has been demonstrated that the observable footprints predicted by
MG theories are strongly degenerate with the signatures induced by the
presence of massive neutrinos. Indeed, the typical
range of the fifth force for $f(R)$ models, determined by the
Compton wavelength $\mu^{-1}$, can reach a few tens of megaparsecs
\citep[see e.g.][]{Cataneo2015} depending on the value of the
parameter $f_{R0}$, and it is comparable with the free-streaming
scale of neutrinos, which can be estimated with
Eq. \eqref{eq:free-streaming_length}. The neutrinos free-streaming
can have thus a counteractive effect on the enhanced growth of the
cosmic structures, causing a compensation on the cosmological
statistical variations given by MG theories. This poses a notable
challenge for cosmology, since robust methods and different
cosmological probes are required to achieve tight constraints on both
massive neutrinos and MG, and especially to disentangle their combined
effects. In this context, cosmic voids can contribute
as key probes, given their peculiar underdense nature and
exceptional spatial extension, comparable to the ranges covered by
the fifth force of $f(R)$ models and by the neutrino free-streaming,
that make them particularly sensitive to both these components.


\subsection {The void size function}
\label{sec:void_size_function_theory}

In the last decade, cosmic voids have demonstrated their potential as
cosmological tools. In particular, void profiles and abundances
constitute some of the most promising statistics to
exploit. While the former has been already exploited
to derive extremely tight cosmological constraints \citep{Hamaus2020, Aubert2020, Hawken2020, Nadathur2020}, the latter, also called the \textit{void size function}\footnote{The terms \textit{void size function}, \textit{void number counts} and \textit{void abundance} are equivalently used throughout the paper to refer to the same observable, that is to the comoving number density of cosmic voids as a function of their size.}, has not been successfully applied to real data yet.

The void size function commonly rely on the theoretical model
developed by \cite{SVdW2004} (hereafter the SvdW model), which is
derived following the same excursion-set approach used to compute the
halo mass function \citep{Press&Schechter, cole1991, Bond1991,
  mo_white1996, ST1999, SMT2001}. Note that, conversely to what
happens during the collapse of overdensities (i.e. DM haloes), the
possible initial non-sphericity of the underdensity perturbations
tends to vanish during the isolated evolution of voids. This suggests
that the adoption of a simple spherical expansion model may be
accurate enough in describing the void formation
\citep{Blumenthal1992}. To find the mathematical expression of the
SvdW model it is required to solve a \textit{two-barrier} problem: one
barrier is necessary to account for void formation and merging
(\textit{voids-in-voids}), and the other for void collapse
(\textit{voids-in-clouds}).  Indeed, for the formation of a void it is
necessary not only to reach a density contrast below a specific
barrier $\delta_v$, but also to avoid being extinguished by a
collapsing overdensity on larger scales, surpassing the threshold
$\delta_c$. The latter is given by the overdensity case, in which a
halo is considered to be formed after its virialisation\footnote{An
alternative value for this threshold is given by the
\textit{turn-around} phenomenon, identifying the moment in which the
overdensity perturbation detaches from the overall expansion of the
Universe. This event occurs in linear theory at a density contrast
$\delta_c = 1.06$ for an EdS Universe. However, since the specific
choice of this value is not relevant for the analysis carried out in
this work, we decide to fix the collapse threshold to the one
related to the halo virialisation.}, that happens when the spherical
perturbation reaches a linear density contrast $\delta_c \approx
1.69$, for an Einstein-de Sitter (EdS) Universe \citep{Bond1991}.  The
barrier $\delta_v$ was instead commonly fixed at the characteristic
value of the \textit{shell-crossing} phenomenon, since this event is
often associated to the void formation \citep{Blumenthal1992,
SVdW2004, Jennings2013}. From the theoretical point of view, with
the excursion-set formalism we consider an initial negative top-hat
perturbation and we model it as a set of concentric
shells. Since the repulsive force experienced by the
internal layers decreases as a function of the radius
\citep{SVdW2004}, the inner shells will expand faster than the
external ones, surpassing them at the specific density contrast given
by $\delta_v \approx -2.71$, in linear theory and for an EdS
Universe. However, this condition strictly depends on the initial
density profile of the underdensity. Considering more physically
motivated density profiles than a top-hat perturbation, the
shell-crossing unlikely occurs in voids, at least at observable scales
\citep{Shandarin2011, Abel2012, Sutter2014a, Hahn2015,
verza2019}. This implies that voids can be modelled
accurately by adopting the linear theory and that any
underdense threshold $\delta_v$ can in principle be selected to
identify the voids (see Section
\ref{sec:finding_and_cleaning}).

From the solution of the two-barrier problem in linear theory, SvdW derived the theoretical expression of the void size function, that is the comoving number density of cosmic voids, $n$, as a function of their effective radius\footnote{In this paper we use the superscripts $\mathrm{L}$ and $\mathrm{NL}$ for the quantities derived in linear and nonlinear theory, respectively. In absence of any superscript, we take for granted the reference to the nonlinear counterpart.}, $r^\mathrm{L}$:
 \begin{equation}
  \label{eq:}
  \frac{\mathrm{d}n^\mathrm{L}}{\mathrm{d}\ln r^\mathrm{L}} = \frac{f_{\ln\sigma} (\sigma)}{V(r^\mathrm{L})} \frac{\mathrm{d}\,\ln\sigma^{-1}}{\mathrm{d}\,\ln r^\mathrm{L}}\ \text{,}
\end{equation}
where $V(r^\mathrm{L})$ is the volume of the spherical fluctuation of radius $r^\mathrm{L}$ and $f_{\ln\sigma}$ is the fraction of fluctuations destined to become voids, as predicted by the excursion-set theory:
\begin{equation}
  \label{eq:SF1}
  f_{\ln\sigma} = 2 \sum_{j=1}^{\infty}j \pi x^2 \sin(j \pi
  \mathcal{D})\exp\biggl[-\frac{(j \pi x)^2}{2}\biggr]\, ,
\end{equation}
where
\begin{equation}
  \label{eq:SF2}
  x \equiv \frac{\mathcal{D}}{|\delta_v^\mathrm{L}|}\,\sigma \, ,
\end{equation}
and
\begin{equation}
  \label{eq:SF3}
  \mathcal{D} \equiv \frac{|\delta_v^\mathrm{L}|}{\delta_c^\mathrm{L} + |\delta_v^\mathrm{L}|}\, .
\end{equation}
In the previous equations, $\sigma$ is the square root of the mass variance, while $\delta_v^\mathrm{L}$ and $\delta_c^\mathrm{L}$ are the \textit{merging} and \textit{collapsing} barriers
previously described, respectively. The quantity $\mathcal{D}$ represents instead the \textit{void-and-cloud} factor and parametrises the impact of halo formation on the evolving population of voids. 
To extend this model to the nonlinear regime, SvdW imposed the conservation of the total number of voids during the transition from linearity to nonlinearity. This condition can be achieved introducing a correction factor ${C \propto (1 + \delta_v^\mathrm{NL})^{-1/3}}$ in the void radius:
\begin{equation}
  \label{eq:SF4}
    \frac{\mathrm{d}\,n}{\mathrm{d}\, \ln r} \biggr|_{\text{SvdW}}
    = \frac{\mathrm{d}\,n}{\mathrm{d}\, \ln (C\,r)}\biggr|_{\text{lin}}\ \text{.}  
\end{equation}
The downside of this assumption is that it implies a fraction of volume occupied by voids which can exceed the total volume of the Universe. To address this issue, \citet{Jennings2013} proposed a {\em volume conserving} model (hereafter the Vdn model), in which the void volume fraction of the Universe is kept fixed in the transition to the nonlinear regime:
\begin{equation}
V(r) \mathrm{d}\,n = V(r^\mathrm{L}) \mathrm{d}\,n^\mathrm{L}  \rvert_{r^\mathrm{L} = r^\mathrm{L}(r)}\, .
\end{equation}
With this further prescription, we can finally write down the expression of the void size function according to the Vdn model:
\begin{equation}
  \label{eq:SF5}
  \frac{\mathrm{d}\,n}{\mathrm{d}\, \ln r}\biggr|_{\text{Vdn}} =
 \frac{f_{\ln \sigma}(\sigma)}{V(r)} \frac{\mathrm{d} \ln \sigma^{-1}}{\mathrm{d}\, \ln r^\mathrm{L}} \biggr\rvert_{r^\mathrm{L} = r^\mathrm{L}(r)}\, .
\end{equation}
This model has been applied in different works \citep{Jennings2013, RonconiMarulli2017, Roncarini2019, contarini2019, verza2019} and its validity in the prediction of void abundances has been largely demonstrated, 
provided that the analysed sample is prepared through a proper selection and reshaping of the underdensities identified by the algorithm of void finding (see Section \ref{sec:finding_and_cleaning}), in order to match the features that characterise the theoretical definition of voids.

According to the Vdn model, voids are defined as spherical and
non-overlapping underdensities, identified in the total matter density
field and characterised by an internal density contrast given by
$\delta_v^\mathrm{L}$. To compare the model predictions with the
measured void abundance it is therefore necessary to model voids
according to the theoretical definition given by the Vdn model (Section \ref{sec:finding_and_cleaning} will be dedicated to
the description of this modelling). Nevertheless, since the theoretical
model of the void size function is formulated in linear theory, the
density contrast used to resize the voids in a nonlinear framework, $\delta_{v}^\mathrm{NL}$,
has to be converted properly to be used in the model. For this
purpose, we exploit the fitting formula provided by
\citet{Bernardeau1994} to convert the underdense threshold:
\begin{equation}
  \label{eq:bernardeau}
  \delta_{v}^\mathrm{L} = \mathcal{C}\, \bigl[1 - (1 + \delta_{v}^\mathrm{NL})^{-1/\mathcal{C}}\bigr] \ , \ \text{with } \mathcal{C}=1.594 \, ,
\end{equation}
which has been demonstrated to be especially accurate for the underdense regions.

Dealing with biased tracers, $\delta_v^\mathrm{NL}$ has to be converted to take into account the effect of the tracer bias on the void density profiles.
\cite{contarini2019} have demonstrated that the relation between the DM density contrast inside cosmic voids, $\delta_{v,\mathrm{DM}}$, and the corresponding threshold value in tracer distribution, $\delta_{v,\mathrm{tr}}$, can be modelled with a linear relation $\mathcal{F}$, depending only on the large scale effective bias $b_\mathrm{eff}$:
\begin{equation}
\label{eq:thr_conversion}
\delta^\mathrm{NL}_{v,\mathrm{DM}} = \frac{\delta^\mathrm{NL}_{v,\mathrm{tr}}}{\mathcal{F}(b_\mathrm{eff})}\, .
\end{equation}
While the meaning and the estimation of the linear function $\mathcal{F}$ will be addressed in Section \ref{sec:halo_abundance}, the computation of the large-scale linear bias $b_\mathrm{eff}$ from the tracer two-point correlation function will be not discussed in this paper, since it is estimated performing the same Bayesian statistical analysis described in details in \textit{Appendix A} of \cite{contarini2019} \citep[see also][]{Marulli2013, Marulli2018}. 

Finally, we underline that not only the void size
function, but also $b_\mathrm{eff}$ depends on the presence of the
fifth force and massive neutrinos. They are indeed both strongly
correlated to the growth of cosmic structures, which is in turn
influenced by modification of gravity and neutrino thermal
free-streaming. Therefore the rescaling of the underdensity
threshold $\delta^\mathrm{NL}_{v,\mathrm{tr}}$ by means of the
large-scale effective bias can lead to degenerate effects on the
resulting void abundance. A rigorous study of the interplay between
these effects on the size function of voids identified using
different types of matter tracers is left to future works.


\section{PREPARATION OF THE MOCK DATA SAMPLES} \label{sec:catalogue_preparation}

In this Section we first present the N-body simulations analysed in this work. Then we describe the method applied to obtain the DM halo samples from DM particle catalogues. In the end we focus on the algorithms of void finding and cleaning.

\subsection{The DUSTGRAIN-\textit{pathfinder} simulations}
\label{sec:dustgrain}
In this work we use a subset of the cosmological N-body simulations
suite called DUSTGRAIN-\textit{pathfinder} (Dark Universe Simulations
to Test GRAvity In the presence of Neutrinos).  These simulations have
been specifically designed with the aim of investigating the
degeneracies between $f(R)$ gravity models and massive neutrinos, and
have been recently exploited in different papers finalised to the
study of possible methods to disentangle these cosmic degeneracies,
that is exploiting weak-lensing \citep{Giocoli2018, Peel2018} and
clustering statistics \citep{Jorge2019}, investigating the massive
haloes' abundance \citep{Hagstotz2019}, the large-scale velocity field
\citep{Hagstotz2019b}, and exploring machine learning techniques
\citep{Peel2019, Merten2019}.  The DUSTGRAIN-\textit{pathfinder}
simulations have been carried out using {\small{MG-GADGET}}, a code
based on an updated version of {\small{GADGET2}} \citep{Springel2005}
developed by \cite{Puchwein2013} to include $f(R)$ gravity
models. This code has then been combined with the particle-based
implementation described in \cite{Viel2010} to include the effects of
massive neutrinos.

The DUSTGRAIN-\textit{pathfinder} simulations follow the evolution of an ensemble of $(2\times)768^3$ particles of DM (and massive neutrinos) within a periodic cosmological box of $750 \ \mathrm{Mpc}/h$ per side. In the reference $\Lambda$CDM simulation (i.e. the one characterised by GR and $m_\nu = 0$) the DM particle mass is equal to $m^p_\mathrm{CDM}=8.1 \times 10^{10} \ \mathrm{M_\odot}/h$ and the gravitational softening is set to $\epsilon_g = 25 \ \mathrm{kpc}/h$, corresponding to about $1/40$ of the mean inter-particle separation. The cosmological parameters assumed in these simulations are consistent with the Planck 2015 constraints (see \cite{PlanckCollaboration2016}) $\Omega_\mathrm{M}=\Omega_\mathrm{CDM}+\Omega_\mathrm{b}+\Omega_\nu=0.31345$, $\Omega_\Lambda=0.68655$, $h=0.6731$, $\mathcal{A}_s=2.199 \times 10^{-9}$, $n_s=0.9658$, which give for the $\Lambda$CDM case an amplitude of linear density fluctuations smoothed on a scale of $8 \ \mathrm{Mpc}/h$ equal to $\sigma_8=0.842$. The remaining set of simulations is created to sample the joint $f(R)-m_\nu$ parameter space. The $|f_{R0}|$ parameter assumes the values in the range $[10^{-6} \text{-} 10^{-4}]$, while $m_\nu$ belongs to the range $[0 \text{-} 0.3] \ \mathrm{eV}$. All the parameters characterising the simulations considered in this paper are reported in Table \ref{tab:models}. Note that the total $\Omega_\mathrm{M}$ (including neutrinos) is kept fixed to compare the density power spectrum between cosmologies with and without neutrinos. This results in equal positions of the peak of the power spectrum and ensures that the spectra are identical in the long-wavelength limit. For a more detailed description of the DUSTGRAIN-\textit{pathfinder} simulations see \cite{Giocoli2018} and \cite{Hagstotz2019}.

\begin{table*}
\caption{Summary of the main numerical and cosmological parameters related to the subset of the DUSTGRAIN-\textit{pathfinder} simulations considered in this work. These simulations are carried out in a volume of $(750 \ \mathrm{Mpc}/h)^3$ and are composed by $768^3$ CDM particles for the $\Lambda$CDM, fR4, fR5 and fR6 models, with the addition of as many massive neutrino particles for the non-$\Lambda$CDM cases. The third column provides the value of the modified gravity parameter $f_{R0}$, while the fourth column the neutrino mass $m_\nu$. The other columns provide $\Omega_\mathrm{CDM}$ and $\Omega_{\nu}$, that are the CDM and neutrino density parameters respectively, and the CDM and neutrino particle masses $m^p_\mathrm{CDM}$ and $m^p_\nu$. The value in the last column is the $\sigma_8$ parameter, which corresponds to the linear power normalisation computed at $z=0$.}
\centering
\begin{tabular}{lcccccccc}
\toprule
Simulation name & Gravity model  &  
$f_{R0} $ &
$m_{\nu }$ [eV] &
$\Omega _{\rm CDM}$ &
$\Omega _{\nu }$ &
$m^{p}_{\rm CDM}$ [M$_{\odot }/h$] &
$m^{p}_{\nu }$ [M$_{\odot }/h$] & $\sigma_8$ \\
\midrule
$\Lambda$CDM & GR & -- & 0 & 0.31345 & 0 & $8.1\times 10^{10}$  & 0 & $0.842$ \\ 
fR4 & $f(R)$  & $-1\times 10^{-4}$ & 0 & 0.31345 & 0 & $8.1\times 10^{10}$  & 0 & $0.963$ \\ 
fR4\_0.3eV & $f(R)$  & $-1\times 10^{-4}$ & 0.3 & 0.30630 & 0.00715 & $7.92\times 10^{10}$ & $1.85\times 10^{9}$ & $0.887$ \\ 
fR5 & $f(R)$  & $-1\times 10^{-5}$ & 0 & 0.31345 &0  & $8.1\times 10^{10}$  & 0 & $0.898$ \\ 
fR5\_0.1eV & $f(R)$  & $-1\times 10^{-5}$ & 0.1 & 0.31107 & 0.00238 & $8.04\times 10^{10}$ & $6.16\times 10^{8}$ & $0.872$ \\ 
fR5\_0.15eV & $f(R)$  & $-1\times 10^{-5}$ & 0.15 & 0.30987 & 0.00358 & $8.01\times 10^{10}$ & $9.25\times 10^{8}$ & $0.859$ \\ 
fR6 & $f(R)$  & $-1\times 10^{-6}$ & 0 & 0.31345 & 0 & $8.1\times 10^{10}$  & 0 & $0.856$ \\ 
fR6\_0.06eV & $f(R)$  & $-1\times 10^{-6}$ & 0.06 & 0.31202 & 0.00143 & $8.07\times 10^{10}$ & $3.7\times 10^{8}$  & $0.842$ \\
fR6\_0.1eV & $f(R)$  & $-1\times 10^{-6}$ & 0.1 & 0.31107 & 0.00238 & $8.04\times 10^{10}$ & $6.16\times 10^{8}$ & $0.831$ \\ 
\hline
\bottomrule
\end{tabular}
\label{tab:models}
\end{table*}

Among all the comoving snapshots available for this project, we select
the ones at the redshifts $z=0, 0.5, 1, 2$, considering only CDM particles also in the case of simulations
containing massive neutrinos, though this assumption
does not have a major impact on the resulting halo catalogues
\citep[see e.g.][]{Villaescusa-Navarro2013b,
Villaescusa-Navarro2014, Castorina2014, Lazeyras2020}. In the
following analyses concerning voids in the DM density fields, we apply
a subsampling factor to DM particles to reduce the computational time,
keeping the $25\%$ of the original particle sample. The collapsed DM
structures have been identified for each snapshot, as described in
\cite{Despali2016}. In particular, the halo catalogues have been
obtained by applying the {\small{Denhf}} algorithm \citep{Tormen2004,
Giocoli2008} to the DM particle sample, finding DM
haloes as gravitationally bound structures, without including
sub-haloes. When running {\small{Denhf}}, we identify virialised
and bound gravitational spherical structures centred on the densest
particles.  Their virial radius, $R_\mathrm{c}$, and viral mass,
$M_\mathrm{c}$, are defined by the relation:
\begin{equation}
  \label{eq:halo_finder}
  \frac{4}{3}\pi R_c^3 \delta_\mathrm{c} \rho_\mathrm{crit} = M_\mathrm{c}\, ,
\end{equation}
where $\rho_\mathrm{crit} \equiv 3H^2/8 \pi G $ represents the
critical density of the Universe and $\delta_\mathrm{c}$ is fixed to
$200$ or $500$. In this work we have employed 200c halo catalogues
only, thus those derived imposing $\delta_c=200$, except for the
comparison test shown in Figure \ref{fig:bias_relations}. The 500c
haloes, identified with $\delta_c=500$, are indeed rarer objects and
their sparsity does not allow to identify a sufficiently large sample
of cosmic voids.  Moreover, we reject the haloes with a number of
embedded DM particles less than $30$, in order to keep only
statistically relevant objects and to avoid contamination by spurious
density fluctuations. This mass cut corresponds to
$M_\textrm{min} = 2.43 \ \mathrm{M_{\odot}}/h$ for the $\Lambda$CDM
case, and has been chosen to select a complete and, at the same
time, dense enough sample of DM haloes, which is fundamental for
identifying a statistically significant number of cosmic voids. The effect of this
assumption has been investigated by repeating the analysis with
different low mass selections, with the requirement of having a good
agreement between the measured effective bias of DM haloes (see
Section \ref{sec:void_size_function_theory}) and the theoretical
predictions, which we compute using the \citet{Tinker2010} model
convolved with the halo mass function of the simulations (see
e.g. Eq. A$6$ in \textit{Appendix A} of \citealt{contarini2019}).

Two of the most important characteristics of the
employed DM halo catalogues are the volume and the spatial
resolution. The sample volume settles indeed the statistical
relevance of the measured void abundance and affects the probability
of finding large voids, while the spatial resolution determines the
smallest scales at which the void number counts are not affected by
numerical incompleteness. It is therefore fundamental to compare
these quantities with those of the upcoming wide field surveys like
the ESA \textit{Euclid}
mission\footnote{\href{http://www.euclid-ec.org}{http://www.euclid-ec.org}}
\citep{Laureijs2011, Amendola2018}, the NASA space mission Wide
Field Infrared Survey Telescope
(WFIRST)\footnote{\href{https://wfirst.gsfc.nasa.gov}{https://wfirst.gsfc.nasa.gov}}
\citep{WFIRST2012} and the Vera C. Rubin Observatory
LSST\footnote{Legacy Survey of Space and Time;
\href{http://www.lsst.org}{http://www.lsst.org}} \citep{LSST2012},
in order to put our results into the context of a future application
with real data catalogues.  At first, with a volume of about $0.42
\ (\mathrm{Gpc}/h)^3$, the DUSTGRAIN-\textit{pathfinder} simulations
allow the identification of a relatively small sample of voids. This
volume can be easily compared to the one of WFIRST, \textit{Euclid}
and LSST, that will cover about $17 \ (\mathrm{Gpc}/h)^3$, $44
\ (\mathrm{Gpc}/h)^3$ and $154 \ (\mathrm{Gpc}/h)^3$,
respectively. Considering these volumes (two of which that are more
than $100$ times larger than the simulation volume considered in
this work), we expect that the uncertainties related to the void
statistics presented in this paper will decrease dramatically
considering void samples extracted from future real galaxy
catalogues. In particular, the Poissonian errors associated to both
the stacked density profiles (see Section \ref{sec:void_profiles})
and void size functions (see Section \ref{sec:halo_abundance}) will
be reduced of a factor proportional to the square root of the
increasing of the volume size, hence allowing to achieve a precision
more than $10$ times better.

Finally, the DM halo catalogues extracted from the
DUSTGRAIN-\textit{pathfinder} $\Lambda$CDM simulations are
characterised by a mean particle separation (MPS) between $8.7$ and
$12.4 \ \mathrm{Mpc}/h$, for $z=0$ and $z=2$ respectively. These
values are indeed of the same order of the expected MPS for the
\textit{Euclid} spectroscopic survey that, with a sky area of $\sim
15000 \ \mathrm{deg}^2$, will sample over $50$ million of H$\alpha$
galaxies, reaching a spatial resolution of about $10
\ \mathrm{Mpc}/h$. The predictions for the spatial resolution of the
WFIRST survey are even more encouraging, thanks to the $20$ million
H$\alpha$ galaxies that are expected to be detected in the redshift
range $1.05<z<2$, sampled over a sky area of $\sim 2400
\ \mathrm{deg}^2$.
Considering instead the LSST photometric galaxy surveys, the MPS of
the samples would drop to $\sim 3 \ \mathrm{Mpc}/h$, though with a
dramatic decreasing of the redshift accuracy.

We can therefore expect that the methodology presented
in this paper will gain statistical relevance when applied to the
data of the upcoming surveys.

\subsection{Void finding and cleaning}
\label{sec:finding_and_cleaning}
In this work we identify cosmic voids by means of {\small
VIDE}\footnote{\url{https://bitbucket.org/cosmicvoids/vide_public}}
\citep{SutterVIDE2015}, a public toolkit which implements an enhanced
version of the void detection code {\small ZOBOV}
\citep{Neyrinck2008}. Belonging to the class of algorithms based on
geometrical criteria, {\small VIDE} searches for local minima in a
three-dimensional density field, reconstructed from the tracer
positions via the \textit{Voronoi tesselation}. Then it groups nearby
Voronoi cells into zones, merging the adjacent ones to form
voids. This procedure makes use of a \textit{watershed} algorithm,
that allows the addition of adjacent zones to a void only if the
density of the ridge between them is less than $0.2$ times the mean
particle density. The resulting catalogue is constituted of a nested
hierarchy of voids, with no assumption about their shapes. Void
centres are defined as the volume-weighted barycentre, $\overline{X}$,
of the $N$ Voronoi cells that define the void:
\begin{equation}
    \overline{X}  = \frac{\sum_{i=1}^{N}{\overline{x}_i V_i}}{\sum_{i=1}^{N}{V_i}} \textrm{ ,}
\end{equation}
where $\overline{x}_i$ are the coordinates of the $i$-th tracer belonging to that void, and $V_i$ the volume of its associated Voronoi cell. The effective void radius, $r_v$, is instead calculated from the total volume of the void, $V_v$, defined as the radius of a sphere having the same volume:
\begin{equation}
    V_v \equiv \sum_{i=1}^{N}{V_i} = \frac{4\pi}{3}r_v^3\textrm{ .}
\end{equation} 
We run {\small VIDE} on both the DM particle and DM halo
distributions, building a void catalogue for each of the cosmological
simulations and redshift considered in this work (see Section \ref{sec:dustgrain}).

\begin{figure*}
\centering
    \includegraphics[width=\textwidth]{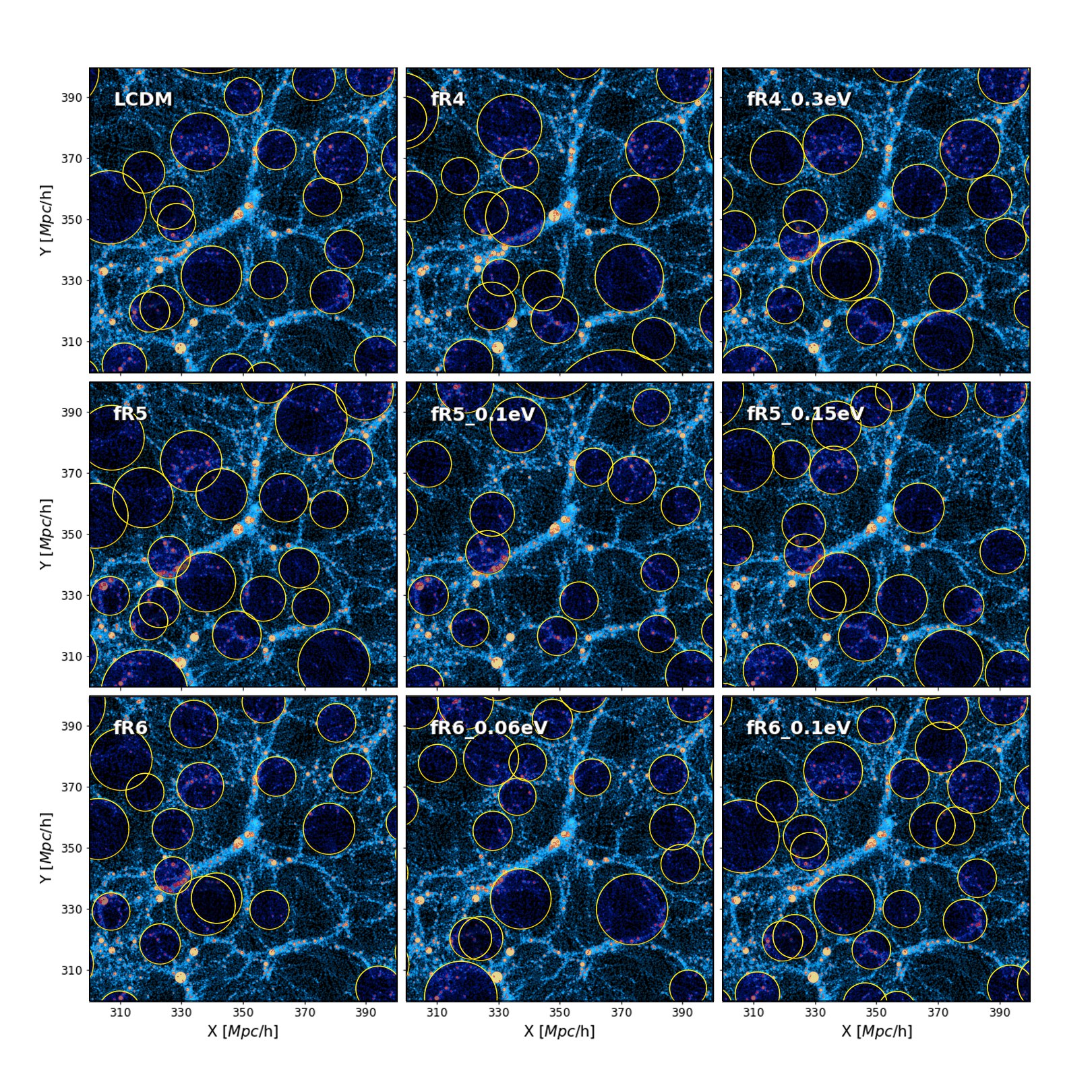}
    \caption{Visual representation of the voids identified in the DM distribution of the DUSTGRAIN-\textit{pathfinder} simulations. We show a slice of $20 \ \mathrm{Mpc}/h$ of the central part of the simulation box at $z=0$, for each of the cosmological scenarios analysed in this work. The DM particles are displayed in light blue, while the DM haloes have colours from orange to yellow, the latter indicating the more massive ones. The yellow circles with a darker interior represent the voids obtained after the application of the cleaner procedure to the void catalogues previously built by applying the {\small VIDE} algorithm.}
    \label{fig:dustgrain_voids} 
\end{figure*}

However, cosmic voids identified by {\small VIDE} do not satisfy all
the assumptions adopted in the theoretical model of the void size
function (see Section \ref{sec:void_size_function_theory}). Indeed,
according to the definition used in this analysis, voids are spherical
non-overlapping regions, centred in density depths of the density
field, embedding a fixed density contrast. Therefore we apply a
cleaning procedure aimed at aligning the objects
included in the void catalogue with the definition of cosmic voids
assumed in the Vdn model. This type of procedure was proposed for
the first time in \citet{Jennings2013}, where it was applied to
void catalogues built with {\small ZOBOV}, to verify the validity of
the methodology using unbiased tracers. In this work, we apply to
the {\small VIDE} void catalogues an improved version\footnote{The
code is included in the {\small CosmoBolognaLib} V5.4
\citep{marulli2016}, a large set of {\em free software} C++/Python
libraries in constant development, available at
\url{https://gitlab.com/federicomarulli/CosmoBolognaLib}. In this
updated version of the cleaning algorithm, we optimised the code by
parallelising its slower parts. We also improved the computation of
the effective void radii by means of a third-order
polynomial fit of the void density profiles, aimed at
reconstructing a smooth trend of the density contrast as a
function of the distance from the void centres, used to rescale
the void radii.} of the algorithm developed by
\cite{RonconiMarulli2017}. As described in the
aforementioned paper, this algorithm is structured in three steps,
designed to select and reshape the detected underdensities to match
the definition used to develop the theoretical model of the void
size function. In particular, in the first step the algorithm
rejects the cases of \textit{voids-in-voids} and
\textit{voids-in-clouds} from the input void catalogue, together with
the underdensities having a radius outside a spatial range selected
by the user. Then, in the second step, it resizes the radius of each
void to a specific value $R_\mathrm{eff}$, enclosing a given value
of the density contrast. Lastly, during the third step the
code checks for pairs of overlapping voids, removing the ones with
the higher central density. All these steps are independent of the
assumed cosmology and of the specific employed void finder. For a more
detailed description see \cite{RonconiMarulli2017, Roncarini2019,
contarini2019}.

At the end of the cleaning procedure, the initial void
catalogue built with {\small VIDE} is pruned of spurious voids, and
consists of properly rescaled underdensities following the definition adopted in the Vdn
model. Therefore our void catalogues turn out to be composed by
non-overlapping spherical objects of radius $R_\mathrm{eff}$,
characterised by an internal density contrast
$\delta^\mathrm{NL}_v$. As previously mentioned, this value can be
fixed to any reasonable threshold, as far as it is both low enough to
identify cosmic depressions and high enough to sample zones spatially
resolved in the matter distribution. Whatever is the value selected to
resize voids, this threshold has to be properly converted using
Eq. \eqref{eq:bernardeau}, with a previous rescaling by means of
Eq. \eqref{eq:thr_conversion} in the case of biased tracers, and then
inserted in Eq. \eqref{eq:SF5} to compute the Vdn model and predict
the void abundance of the sample. With this
prescription, the theoretical model will be set up to predict the
abundance of voids with different depth, according to the threshold
selected to rescale the underdensities. However, even if the
agreement between the theoretical and the measured void size
function is kept for every selection of the threshold
$\delta_v^\mathrm{NL}$, the statistical relevance of the results can
depend on this choice. Indeed, choosing a particularly low
underdensity threshold (e.g. $\delta_v^\mathrm{NL}=-0.9$), the
scaling algorithm will rescale void towards very small radii, in
order to reach this spherical density contrast. This rescaling will
cause the leak of void counts due to the increase of small voids
that will be discarded during the subsequent exclusion of
underdensities with radius belonging to poorly sampled regions,
selected according to the MPS of the tracers. Analogously,
choosing an underdensity threshold particularly high
(e.g. $\delta_v^\mathrm{NL}=-0.5$) the final sample will be composed
by larger rescaled voids, that will be rejected more frequently
during the cleaning procedure because of their greater propensity of
overlapping, causing also in this case the decreasing of void number
counts.

To clean the catalogues of voids identified in the DM halo
distribution, we fix the threshold $\delta^\mathrm{NL}_{v,
\mathrm{tr}}$ at the value $-0.7$, following the choice made in
\cite{contarini2019}. Indeed, although voids are shallower depressions
at earlier cosmic times, the bias dependence in the
$\delta^\mathrm{NL}_{v, \mathrm{DM}}$ threshold makes the values
high enough to be reached also by voids at high redshifts.  On the
contrary, dealing with voids in the DM particle distribution, the
value $\delta^\mathrm{NL}_{v, \mathrm{DM}}=-0.7$ is less appropriate
to identify cosmic underdensities. Indeed, only few and very deep
voids could be rescaled to enclose such a low density contrast at high
redshifts. Since the choice of the threshold does not affect the
validity of the predictions of the Vdn model, we decided to use higher
density contrasts to clean voids at earlier epochs,
adopting different thresholds depending on the
redshift of the DM catalogues, using the growth factor $D$ to
rescale the nonlinear density contrast required in the cleaning
algorithm:
\begin{equation}
  \delta_{v, \mathrm{DM}} (z) = -0.8 \, \bigg( \frac{D(z)}{D(z=0)}
  \bigg)^2 \textrm{ .}
\end{equation} 
Fixing the cosmological parameters to those of the $\Lambda$CDM
simulations, we obtain the following values: ${\delta_{v,
    \mathrm{DM}}(z=0)}=-0.80$, ${\delta_{v,
    \mathrm{DM}}(z=0.5)}=-0.70$, ${\delta_{v,
    \mathrm{DM}}(z=1)}=-0.62$ and ${\delta_{v,
    \mathrm{DM}}(z=2)}=-0.52$.  It is important to highlight, once
again, that the matching between the measured void abundance and the
predictions of the Vdn model is not affected by the specific choice of
the underdensity threshold. As far as the same value is used to
reshape voids and is also inserted, after the conversion in its linear
counterpart, in Eq. \eqref{eq:SF5}, the results will be in agreement
with the model predictions. Therefore the reader should not be misled
by the fact that the cleaning procedure is, in this
case, cosmology dependent. The usage of the growth factor is just a
convenient prescription to select an effective threshold, depending on
the redshift of the sample, and does not introduce any
cosmology-driven bias. The increasing of the
underdensity threshold with the redshift is performed to enlarge the
sample of voids and reduce the shot noise. However, it is important
to point out that this choice can affect the purity of the void
catalogue since these voids are characterised by shallower internal
density contrasts and are therefore more prone to be
spurious\footnote{Effective techniques to evaluate
the purity of a void catalogue and reduce the background
contamination are presented e.g. in \citet{Neyrinck2008} and
\citet{Cousinou2019}.}. We also tested
different values of the threshold, finding consistent outcomes, as
it also has been proved in previous works \citep{Roncarini2019,
contarini2019, verza2019}. Among the tested thresholds, we
selected the most effective ones to maximise the signal and reduce
the noise associated to the measured void abundances. In particular,
we aimed at obtaining a total number of voids that was as large as
possible, taking into account the spatial limit given by the
numerical incompleteness affecting the void size function (see also
Sections \ref{sec:DM_abundance} and \ref{sec:halo_abundance}).

Figure \ref{fig:dustgrain_voids} shows the voids identified in the
distribution of DM particles at $z=0$, obtained following the cleaning
procedure described in this section. For each cosmological model, we
report the central regions of the simulation box, indicating the
spherical underdensities selected in this work with circles traced
within a slice of $20 \ \mathrm{Mpc}/h$ along the Z-axis. Any apparent
overlapping between voids is a visual effect caused by the projection
on the plane. As expected, the denser zones made up by filaments are
rejected from the selection. Some empty regions result not identified
as voids, due to the superimposition with other underdensities not
displayed in the figure. It is also interesting to note that the
selected sample of voids is different depending on the cosmological
scenario, even if the underlying distribution of matter looks
remarkably similar, which does not depend on the
cleaning procedure, as we verified.


\begin{figure*}
\centering
    \includegraphics[width=0.47\textwidth]{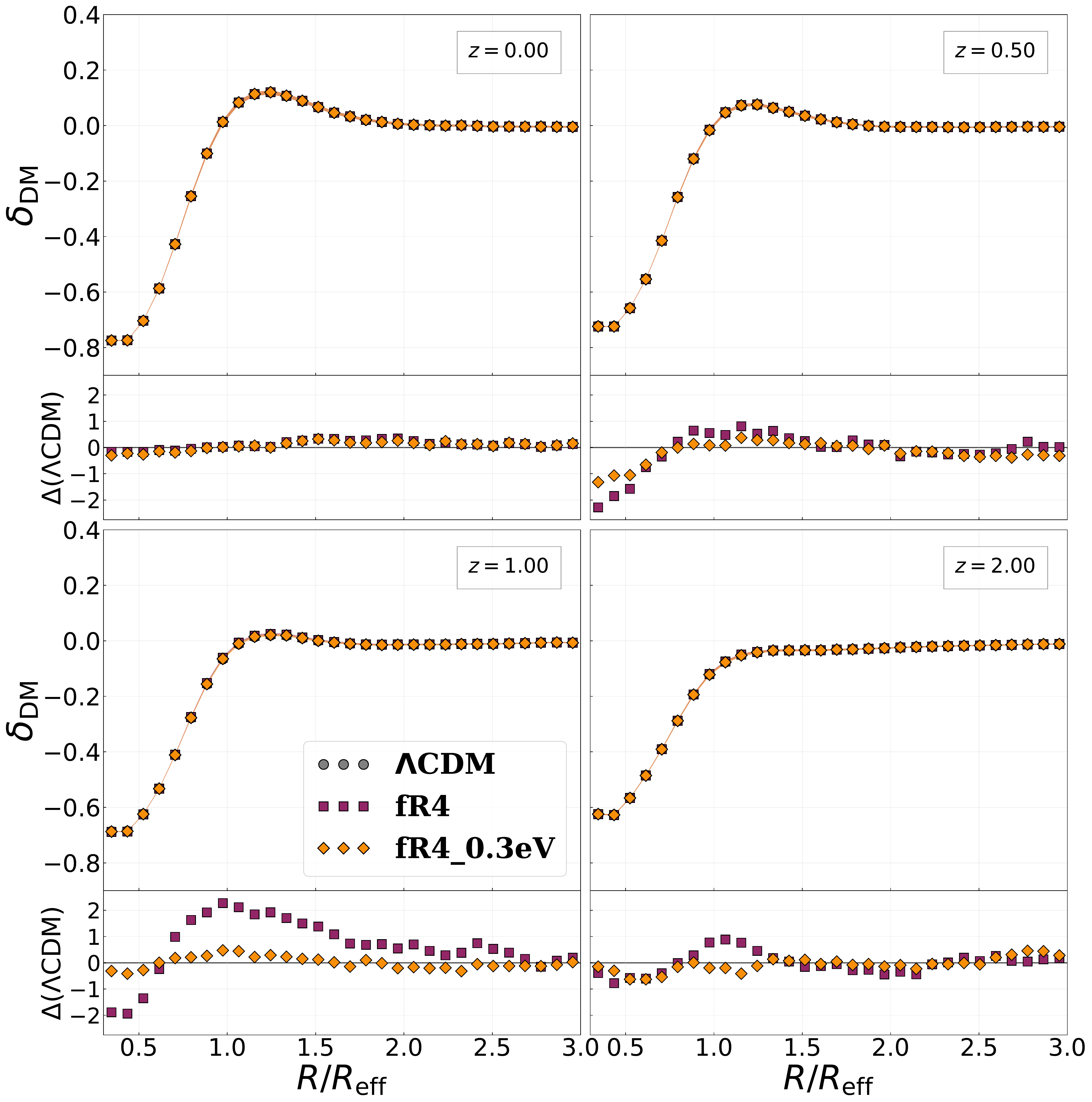}
    \includegraphics[width=0.267\textwidth]{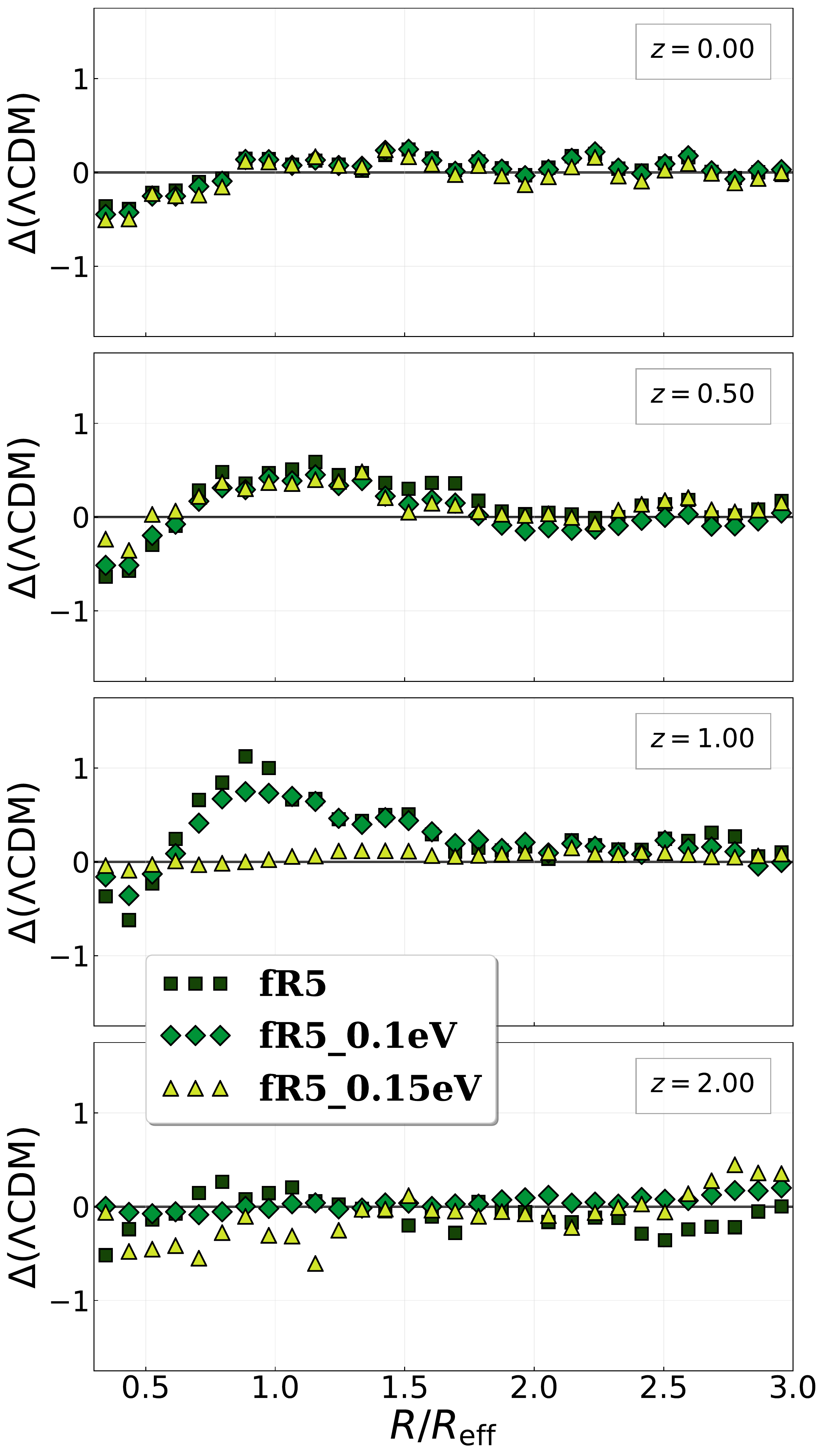}
    \includegraphics[width=0.254\textwidth]{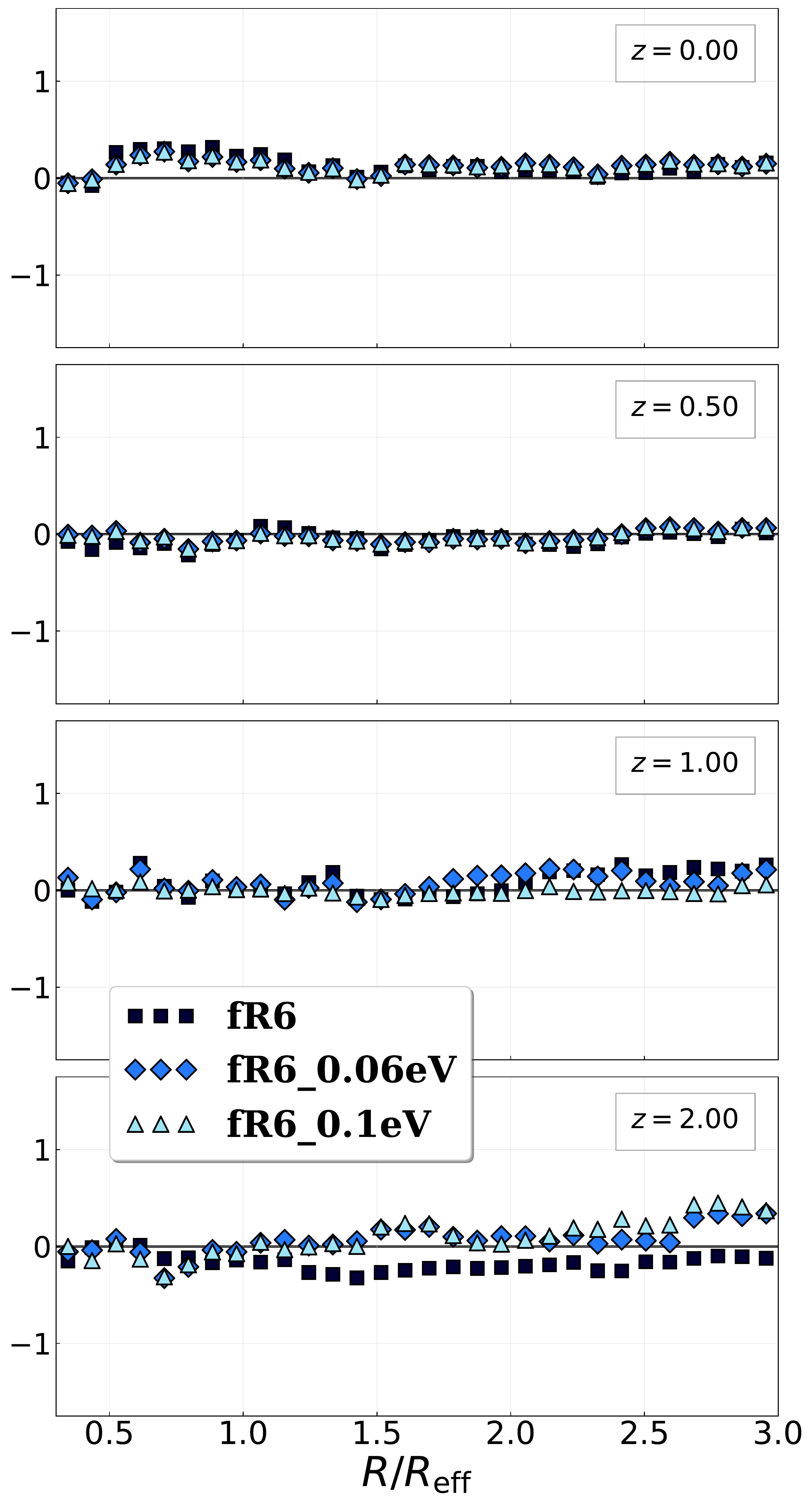}
    \caption{Density contrast profiles computed in shells around the centres of
      cosmic voids, identified with {\small VIDE} in the distribution
      of DM particles. The results are displayed for each cosmological
      model at redshifts $z=0, 0.5, 1, 2$. We report in the left plot
      the profiles measured considering the $\Lambda$CDM, fR4 and fR4\_0.3eV
      simulations. These profiles are so similar that the markers 
      with which they are represented result superimposed. However, the differences
      between them are highlighted in the residuals reported in each sub-panel,
      computed with respect to the $\Lambda$CDM case, in units of the errors associated to the profiles computed in the \textit{non-standard} cosmologies. 
      The latter are represented as a shaded region in the plots. In this case,
      given the high number of profiles, this uncertainty is so small to be represented 
      with a simple line between the data points. To highlight the
      deviations from the $\Lambda$CDM void profiles, we show in the
      sub-panels in the two most-right columns only the residuals
      obtained with the other models
      analysed in this work: fR5 and fR6 MG models, with and without
      massive neutrinos.}
    \label{fig:void_profiles_DM_fR4-5-6} 
\end{figure*}

\section{Results}
\label{sec:results}

In this Section we present the main results obtained in this
work. We first show the void density profiles traced by the DM
particles and by the DM haloes, analysing the differences emerging
between the profiles computed in $f(R)$ gravity and massive neutrino
cosmologies and the ones of the $\Lambda$CDM case. Then we focus our
analysis on the study of the void size function. We compare the
measured void abundance with the theoretical predictions of the Vdn
model in different cosmologies, using voids identified both in the DM
and in the biased tracer distribution, searching for the most
efficient methodology to disentangle the cosmic degeneracies between
MG and neutrino effects.

\subsection{Void profiles}
\label{sec:void_profiles}

Void profiles have been analysed in different works and their study
has demonstrated not only to be promising to derive cosmological
constraints \citep[see e.g.][]{Paz2013, Ricciardelli2014, Pisani2014,
Nadathur2015b, Nadathur2016, Hamaus2020, Aubert2020}, but also to be
useful to understand the overall assembly of the cosmic structures in
filaments and walls \citep{Cai2014, Hamaus2014, Padilla2016,
Massara2018}. We compute the stacked void density profiles by
measuring the density contrast in shells around void centres. In
particular, we calculate the mean of the density profiles computed
between $0.3$ and $3$ times the effective radius $R_\mathrm{eff}$,
rescaling then each profile by its correspondent void effective
radius. For this specific analysis we make use of the void catalogues
obtained directly with {\small VIDE}, without applying the cleaning
algorithm described in Section \ref{sec:finding_and_cleaning}. This is
due to the fact that the cleaning procedure is aimed at shaping voids
according to the theoretical model of the void size function, and it
is not particularly suitable for the study of the stacked void
profiles. Indeed, using our cleaning prescriptions, the sample of
voids is considerably reduced in number because of the removal of the
\textit{voids-in-voids}, \textit{voids-in-clouds} and of the
overlapping cases. Moreover, with the cleaning algorithm we rescale
the void radii to match a specific density contrast, whereas in the
study of the stacked density profiles we aim at modelling voids to
enhance the self-similarity between their shapes. Indeed, the {\small
VIDE} void catalogues are composed by a hierarchy of voids separated
by high density walls, and the effective radius assigned to each void
is, by construction, in proximity to the so-called
\textit{compensation wall}. These voids are therefore characterised by
the same shape and their stacking allows to sharpen their peculiar
features, as the progressive emptying of the underdense internal parts
and the formation of the compensation wall over the cosmic time.

\begin{figure*}
\centering
    \includegraphics[width=0.467\textwidth]{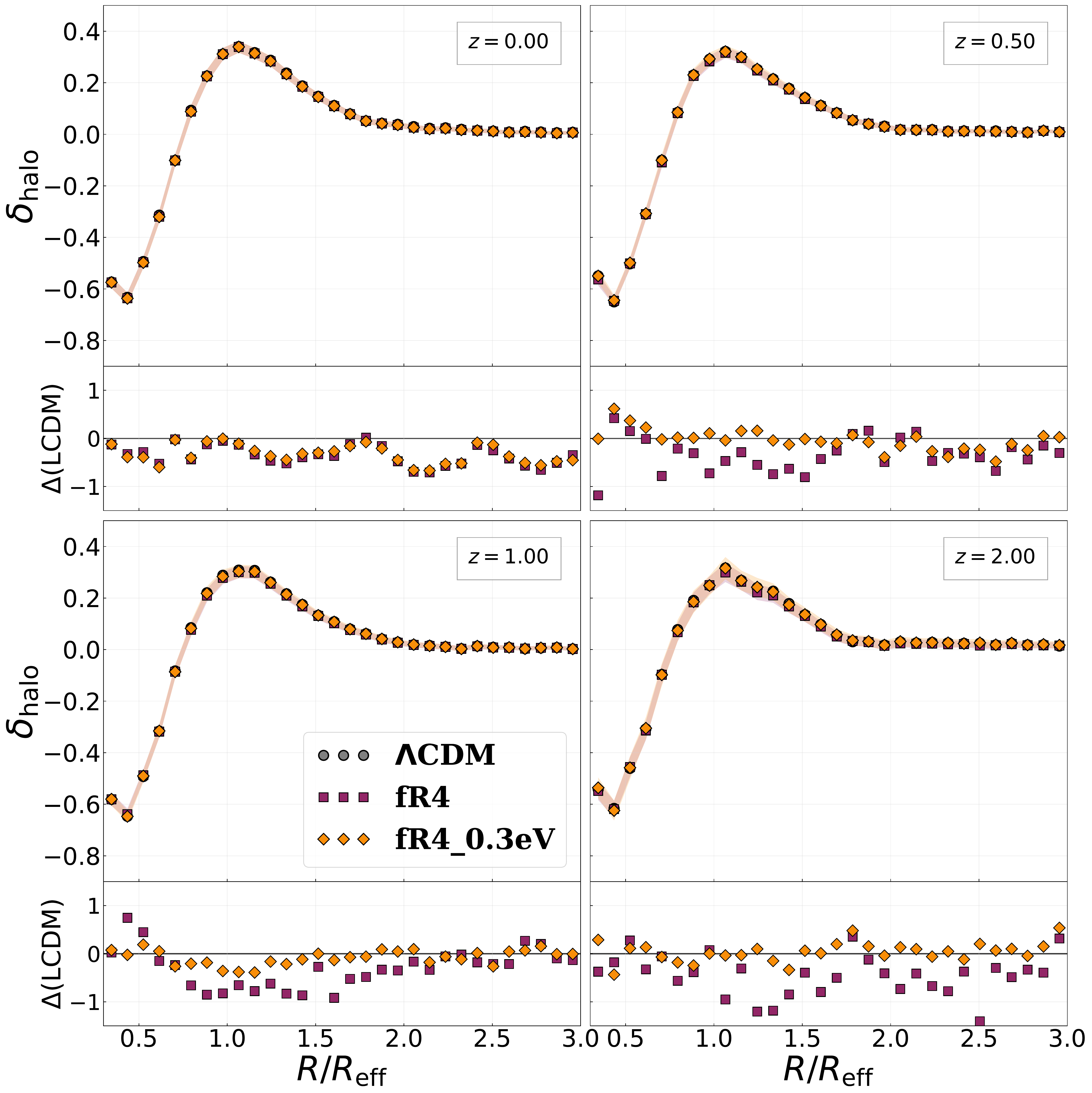}
    \includegraphics[width=0.266\textwidth]{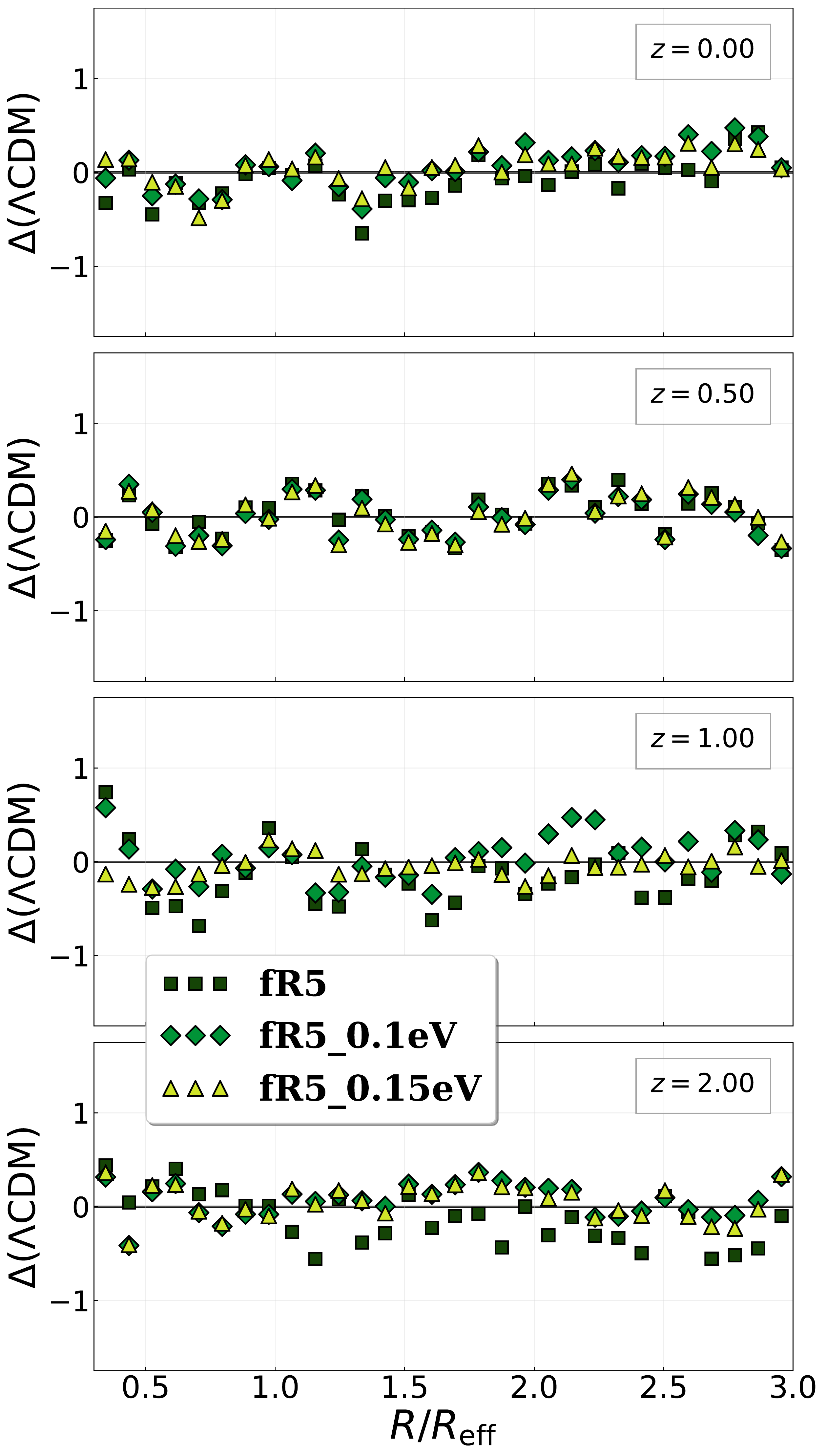}
    \includegraphics[width=0.253\textwidth]{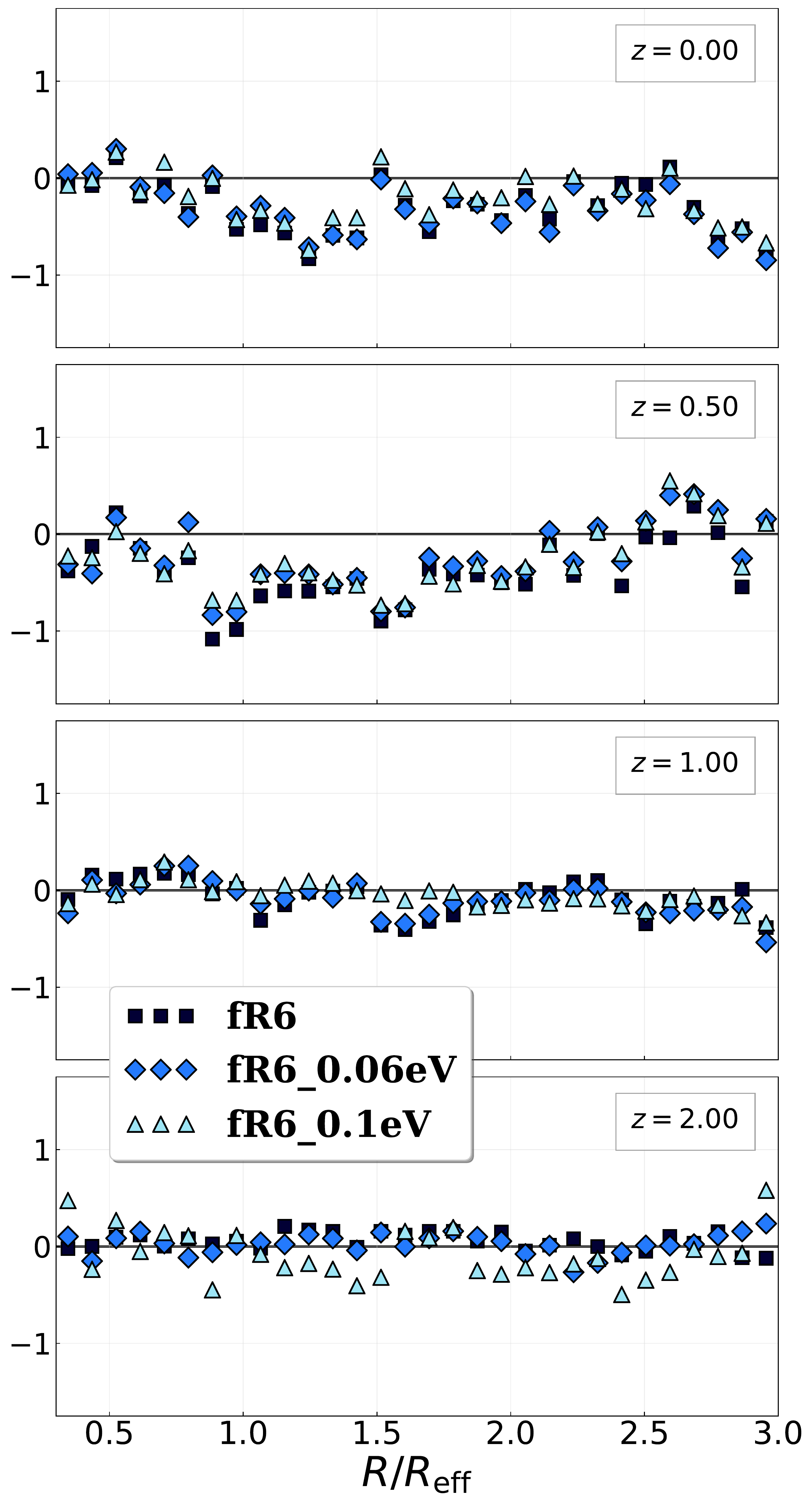}
    \caption{As Figure \ref{fig:void_profiles_DM_fR4-5-6}, but for the cosmic voids identified with {\small VIDE} in the distribution of DM haloes with $\delta_c=200c$.}
    \label{fig:void_profiles_halo_fR4-5-6} 
\end{figure*}

We start analysing the profiles computed in the DM particle
distribution, considering only voids with radii included in the range
$[5 \text{-} 7]$ times the MPS, which corresponds to $1.55 \ \mathrm{Mpc}/h$, for all the sub-sampled catalogues. This
range covers the central parts of the interval on which we perform the
analysis of the abundance of voids in the DM density field presented in the following Section
\ref{sec:DM_abundance}: the lower limit is given by the spatial
resolution of the sample, while the upper limit is chosen to include a
sufficient number of voids with large radii. Since the shape of the
density profiles depends on the mean radius of the void sample
\citep{Hamaus2014}, we avoid to select a wider range of sizes to
prevent an excessive mixing of different density profiles during the
operation of average.  

In Figure \ref{fig:void_profiles_DM_fR4-5-6} we show the stacked
profiles of voids in the DM field at different redshifts, for the $9$
cosmological models considered in this analysis. In the left plot we
report the results obtained with the $\Lambda$CDM simulations,
compared to those with MG models characterised by $f_{R0}=-10^{-4}$,
with and without massive neutrinos with $m_\nu = 0.3 \ \mathrm{eV}$
(namely, fR4 and fR4\_0.3eV). The density profiles in different
cosmologies appear very similar, at all redshifts. We note that the
central zones become deeper with cosmic time, while the compensation
wall grows and turns denser, as already verified in
different works \citep[see e.g.][]{Hamaus2014, Massara2015,
Pollina2016}. Differences among the cosmological models can be
better appreciated by looking at the residuals, displayed in the lower
sub-panels. Here we compute the difference between the mean density
contrast measured in the fR4 or fR4\_0.3eV simulations and the one
measured in the $\Lambda$CDM simulations, divided by the errors
associated to the former. The errors are evaluated as the standard
deviation of all the profiles considered for each simulation, divided
by the square root of their number.  The most significant variation
arises around $z=1$, where the fR4 model shows an increase of the mean
density in close proximity to the compensation wall and a lowering
near the void centres. This is in agreement with the expected effect
of enhancing the growth of structures in MG, that accelerates the
process of void formation and evolution. The presence
of emptier voids and steeper voids profiles has indeed already been
observed and predicted by different authors who studied the behaviour
of the fifth force in voids in Chameleon models \citep[see e.g.][]{
Martino2009, Clampitt2013, Perico2019}. Nevertheless, these
differences are almost completely cancelled by the effect of the
neutrino thermal free-streaming, that slows the
evolution of voids and smooths the void density profiles \citep[see
also][]{Massara2015}, nullifying the possibility of disentangling
the degeneracy between these models. In the right panels of Figure
\ref{fig:void_profiles_DM_fR4-5-6} we present the normalised residuals
obtained by comparing the density profiles measured using the
remaining models to the ones of the $\Lambda$CDM simulation. Note that
the y-range is shrunk compared to the previous plot for the sake of
clarity. Also in the case of fR5 models, the most evident deviations
from the $\Lambda$CDM profiles appear around $z=1$, and they also tend
to vanish in the presence of massive neutrinos. The effect is even
milder in fR6 models, and statistically indistinguishable, at least
with the current simulations.

In order to investigate possible trends related to the void mean size,
we repeated the same analysis dividing the stacked void profiles
into different bins of effective radii. We do not report the results
of this analysis since we did not find any clear 
different behaviour in the
profiles computed with the $\Lambda$CDM cosmology compared to the
other models, at the same mean radii. Minor differences appear only for
voids with large radii at $z=2$, where an early formation of the
compensation wall is revealed in the profiles measured in MG
simulations without massive neutrinos. Larger voids manifest also a
slightly deeper profiles at ${z=0}$ in the very central regions of the voids,
which is reduced by the presence of massive
neutrinos. These results are not surprising given that larger voids
are subject to a faster evolution compared to the smaller
ones. Nevertheless, these deviations do not show a significance higher
than $2 \sigma$, for all the redshifts and distances from the
void centres considered.

\begin{figure}
\centering
\includegraphics[width=1.0\columnwidth]{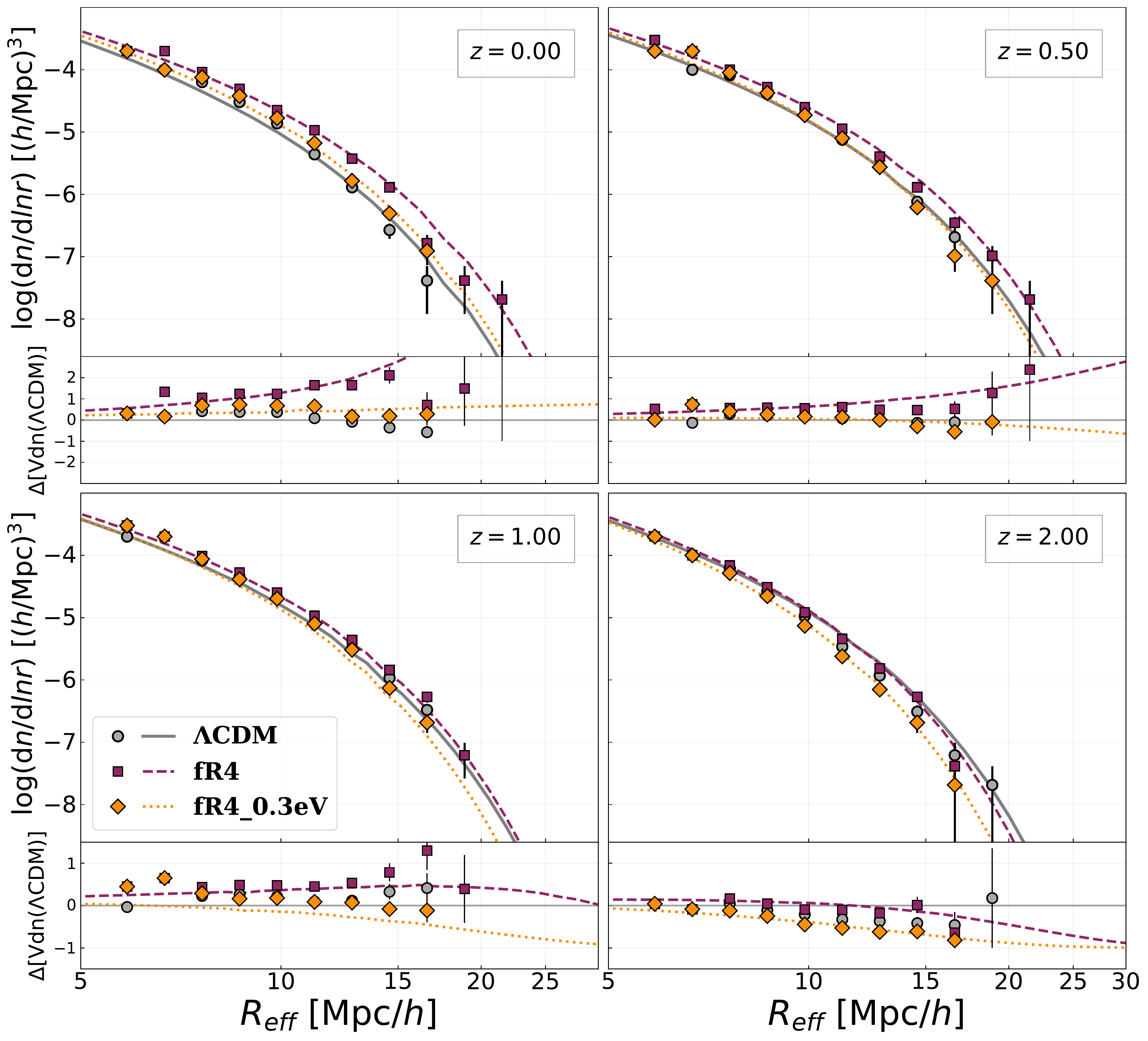}
    \caption{Measured and theoretical void size function computed for
      the $\Lambda$CDM, fR4 and fR4\_0.3eV models, at redshifts $z=0,
      0.5, 1, 2$. The measured void abundances for each cosmological
      model are represented by different markers and colours, as
      described in the label. The errorbars are Poissonian uncertainties
      on the void counts. The predictions are instead displayed as
      lines with different colours and styles, according to the model
      to which they refer. In the bottom sub-panel of each of the $4$
      plots are reported the residuals calculated as the difference
      from the $\Lambda$CDM Vdn model divided by the value
      of the latter, for both the measured and the predicted void abundance.}
    \label{fig:size_functions_DM_fR4} 
\end{figure}

\begin{figure*}
\centering
    \includegraphics[width=1.02\columnwidth]{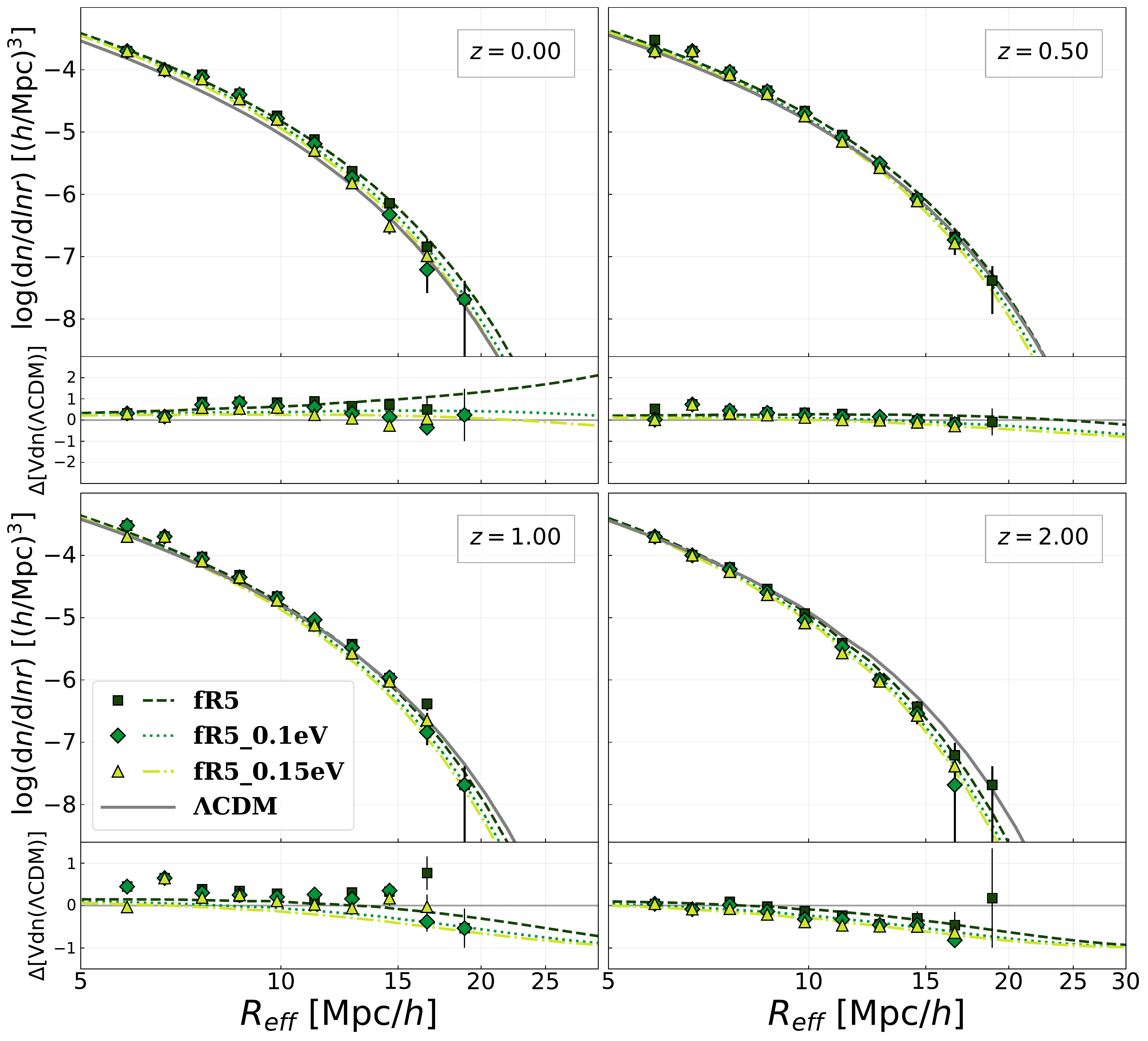}
    \includegraphics[width=1.02\columnwidth]{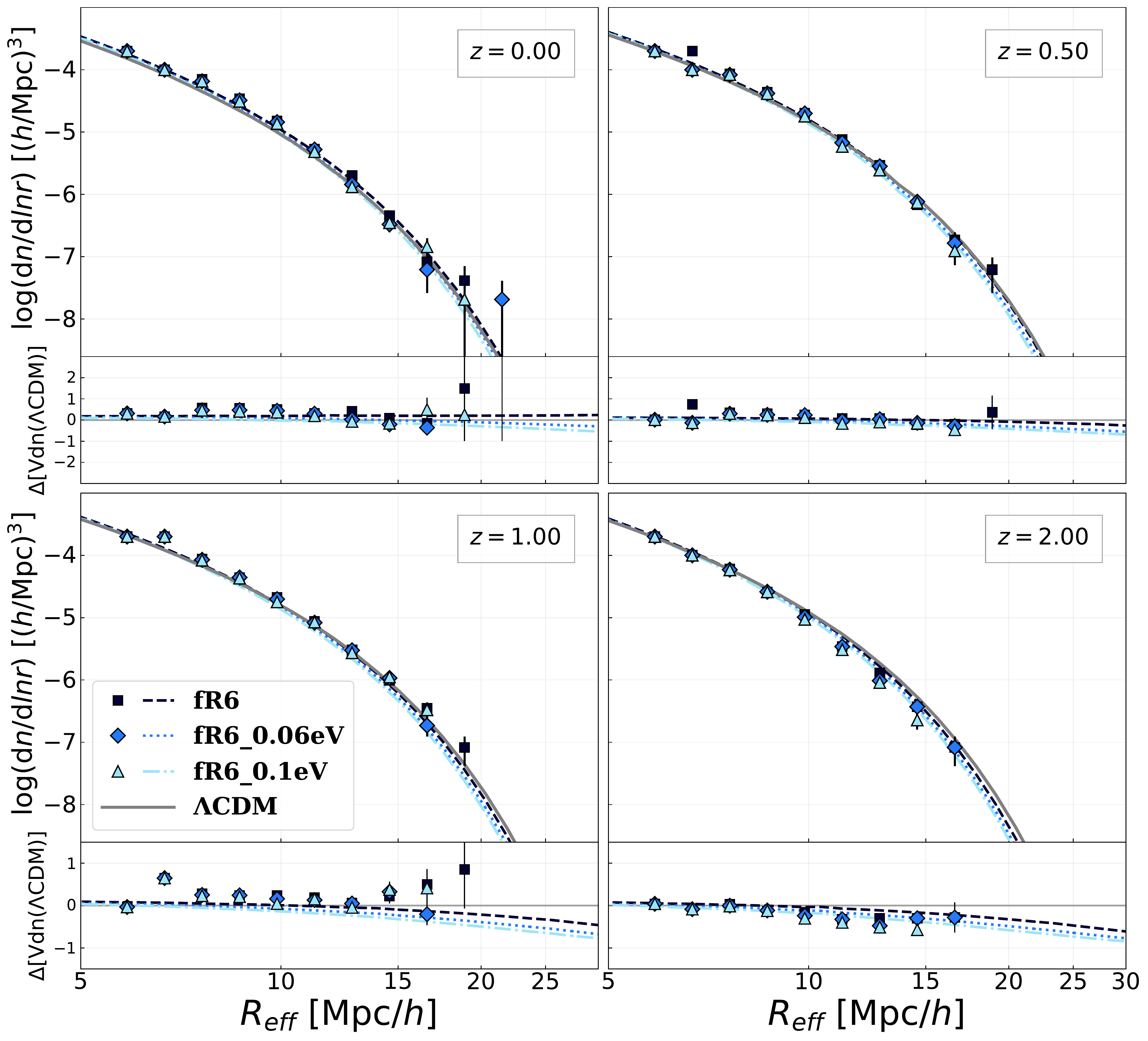}
    \caption{Measured and theoretical void size function computed at redshifts $z=0, 0.5, 1, 2$ for the models fR5, fR5\_0.1eV, fR5\_0.15eV (\textit{left panels}) and fR6, fR6\_0.06eV, fR6\_0.1eV (\textit{right panels}). The description of these plots is analogous to the one reported in the caption of Figure \ref{fig:size_functions_DM_fR4}.}
    \label{fig:size_functions_DM_fR5-6} 
\end{figure*}

Now we present the same analysis performed on void density profiles
measured in the distribution of DM haloes with $\delta_c=200$. In this
case we consider voids with radii in the range $[2 \text{-} 5] \times
\mathrm{MPS}$ of the $\Lambda$CDM simulation tracers, with
$\mathrm{MPS}=8.67 \ \mathrm{Mpc}/h$. The choice of this interval of
radii is motivated by the same reasons behind the previous analysis of
DM void profiles. We report the results of this analysis in Figure
\ref{fig:void_profiles_halo_fR4-5-6}. In the left panel we present the
density profiles for the $\Lambda$CDM, fR4 and fR4\_0.3eV models,
while in the right panels we show the residuals computed for the set
of 6 simulations of the fR5 and fR6 models, with and without massive
neutrinos.  The residuals are computed as the difference between the
profiles measured in \textit{non-standard} cosmological models and the
ones measured in the $\Lambda$CDM cosmology, divided by the
uncertainty associated to the former.  Compared to the
stacked density profiles traced by DM particles, the profiles
obtained using DM haloes result steeper and the compensation wall is
clearly well developed also at early epochs, reaching more positive
values of density contrast \citep[in agreement with the results obtained by][]{Massara2015}.
However, in this case the data are
noisier because of the decreasing of void statistics,
and it is hard to distinguish any significant trend.  Since we expect
to find the strongest deviations in the most extreme MG models, we
focus on the density profiles computed using the simulations with
${f_{R0}=-10^{-4}}$ and $m_\nu = 0.3 \ \mathrm{eV}$. Looking at the
residuals shown in the left panel of Figure
\ref{fig:void_profiles_halo_fR4-5-6} it is possible to note a slight
trend of the fR4 profiles towards lower values of the density
contrast, which is almost completely cancelled by the effect of
massive neutrinos, especially at high redshifts. The origin of these
deviations is the shift of the mean radii of voids identified in MG
scenarios by biased tracers. Indeed, being these voids more evolved
due to the effect of the enhanced gravity, their average radii result
larger. In turn, as demonstrated by \cite{Hamaus2014}, the density
profiles computed with larger voids have shallower interiors and lower
density contrast values in the outer parts. We also tested the
subdivision of the sample in different bins of void radii, but the
increase of the noise does not allow us to discern any characteristic
behaviour associated with voids of different sizes. We can conclude
that the degeneracies between the considered models
cannot be disentangled by the analysis of the void stacked profiles
carried on in this paper, especially making use of DM
haloes as tracers of the matter
distribution. Nevertheless, we underline that larger
simulations could lead to slightly different results, since they would
provide better void statistics and smaller errors, possibly allowing
us to disentangle the models.


\subsection{Void abundance in the DM field}
\label{sec:DM_abundance}

Equipped with the theoretical models delineated in Section
\ref{sec:void_size_function_theory}, we can now focus on the study of
the abundance of cosmic voids as a function of their effective radius.
In this analysis we compare the measured void size function with the
predictions of the Vdn model, making use of samples of voids
identified in the DM particle distribution. The
primary goal of this study is the validation of the methodology with
unbiased tracers, which allows us to perform a precise comparison
between data and models, thanks to the statistically large sample of
detected voids and the absence of the uncertainties related to the
measure of the tracer bias. The application of the method to real
data would be currently infeasible due to the difficulty of
reconstructing the DM density field and the low capability of
surveys to detect a population of voids with sizes comparable to the
ones considered in the following analysis.

We use the simulations with different $f_{R0}$ parameters and neutrino
masses to build the catalogues of voids, exploiting the cleaning
algorithm described in Section \ref{sec:finding_and_cleaning}. To
minimise the effect due to the spatial resolution of the simulations,
we apply the conservative choice of rejecting voids with radii smaller
than $5.5 \ \mathrm{Mpc}/h$, corresponding to about $3.5 \times
\mathrm{MPS}$. When dealing with voids traced by the DM distribution,
no bias prescription is required to rescale void radii. The Vdn model
has indeed been demonstrated to successfully predict the abundance of
voids identified using unbiased matter distributions in standard
$\Lambda$CDM scenarios \citep{Jennings2013, RonconiMarulli2017,
  Roncarini2019}. To include in the theoretical model the variations
caused by MG and massive neutrinos on the void size function, we make
use of {\small
  {MGCAMB}}\footnote{\url{https://github.com/sfu-cosmo/MGCAMB}}
\citep{MGCAMB2009, MGCAMB2011, MGCAMB2019}, a modified version of the
public Einstein-Boltzmann solver {\small CAMB} \citep{CAMB}, which
computes the linear power spectrum for a number of alternative
cosmological scenarios, including the Hu \& Sawicki $f(R)$
model. Since we need to rescale the mass variance at different
redshifts, we have to multiply its value at $z=0$ by the normalised
growth factor $D(z)/D(z=0)$. We derive the latter by means of {\small
  MGCAMB}, computing the square root of the power spectra ratio
$P(z)/P(z=0)$, evaluated on the scales of interest for our analysis.

In Figure \ref{fig:size_functions_DM_fR4} we show the results for the
$\Lambda$CDM, fR4 and fR4\_0.3eV models at redshift $z=0, 0.5, 1,
2$. The measured void abundance is represented by different colours
and markers for each cosmology, while the corresponding Vdn model is
indicated by a line of the same colour. The overall trend of the void
size functions measured in the simulations is well reproduced by the
models. We considered Poissonian errors, thus the uncertainty on the
void counts might be slightly underestimated. In the bottom panels we
report the residuals evaluated with respect to the Vdn model computed
for $\Lambda$CDM case. In particular, the residuals are calculated as
the difference between the measured void abundance and
the corresponding predicted one for a given model and the
theoretical value of the $\Lambda$CDM void size function at the same
radius, divided by the latter. As expected, at low redshifts the fR4
model predicts a larger number of voids with larger sizes. The
modification of gravity induces indeed a faster formation and
evolution of cosmic structures, including cosmic voids. Figure
\ref{fig:size_functions_DM_fR5-6} shows the results of the analysis
performed for the remaining cosmological models, given by the set of
simulations with ${f_{R0}=-10^{-5}}$ and ${f_{R0}=-10^{-6}}$. Also in
these cases, the predictions of the Vdn model computed with {\small
MGCAMG} are fully consistent with the measured void abundance. The
deviations from the $\Lambda$CDM model are weaker in these cases,
given the lower values of the $f_{R0}$ parameter. As
expected, the departure from the $\Lambda$CDM model are the more
severe the stronger is the intensity of the fifth force, resulting
more evident for large voids, in agreement to what found by
\citet{Clampitt2013} and \citet{Voivodic2017}.  It is interesting
to note that, despite at low redshifts the effect of massive neutrinos
is effective in bringing the void size function towards the one
computed in $\Lambda$CDM, this trend starts to revert at higher
redshifts. In particular, it is evident that at $z=2$ the presence of
massive neutrinos makes the fR4\_0.3eV void size function to depart
from the $\Lambda$CDM one, causing a weakening of the growth of
structures, more evident for voids with larger radii. This is a clear
hint of the possibility of disentangling the degeneracies between the
standard $\Lambda$CDM cosmology and MG with massive neutrino
models. However, to achieve this task, it is required to explore the
void abundance at high redshifts and in wide areas, in order to
collect a sufficiently high number of large voids.


\subsection{Void abundance in biased tracer field}
\label{sec:halo_abundance}

The study of the size function of voids identified in a biased tracer
distribution is certainly a fundamental step towards its future
cosmological exploitation. Many efforts have been made in recent years
to understand the effect of a biasing factor on the modelling of cosmic
voids \citep[see e.g.][]{Furlanetto2006, Sutter2014b, Nadathur2015c,
  Roncarini2019, verza2019}. As anticipated in Section
\ref{sec:void_size_function_theory}, \cite{contarini2019} have
introduced a parametrisation of the threshold $\delta_v$ of the Vdn
model to properly take into account the variations on the void
abundance caused by the usage of biased tracers to define voids. The
function $\mathcal{F}$, introduced in Eq. \eqref{eq:thr_conversion}, is used to
convert the tracer bias computed on large scales, $b_\mathrm{eff}$, to
its corresponding value computed inside cosmic voids. Since the
excursion-set theory considers voids identified in the DM
distribution, we need to expand the radii of the spherical voids
predicted by the Vdn model in order to reach the same density contrast
fixed during the cleaning procedure (see Section
\ref{sec:finding_and_cleaning}). To do this, we have to follow the
spherical density profiles of voids and search for the multiplicative
factor required to convert the density contrast computed in the DM
field to the one measured in the biased tracer distribution, at the
\textit{punctual} distance $R_\mathrm{eff}$ from the void
centres. This is the same technique described in details in
\cite{contarini2019}, aimed at computing the value of
$b_\mathrm{punct}$, defined therefore as
\begin{equation}
     b_\mathrm{punct} \equiv \frac{\delta_{v, \mathrm{tr}}(R=R_\mathrm{eff})}{\delta_{v, \mathrm{DM}}(R=R_\mathrm{eff})} \textrm{ .}
\end{equation} 
The latter is measured using the mean spherical density profiles and
represents the bias of the tracers inside cosmic underdensities. Being
this value hardly obtainable in real data catalogues, a conversion
is required to calculate it from the measure of $b_\mathrm{eff}$. In
\cite{contarini2019} a linear function $\mathcal{F}(b_\mathrm{eff})$ has been
calibrated by fitting the values of $b_\mathrm{eff}$ and
$b_\mathrm{punct}$ computed at different redshifts, using Friends-of-Friends (FoF) halo
catalogues extracted from the {\small CoDECS} simulations
\citep{baldi2012}. In this work we apply the same procedure using the
catalogues described in Section \ref{sec:dustgrain}, in particular
those characterised by the $\Lambda$CDM cosmology. We consider both the
halo catalogues obtained by applying the {\small Denhf} algorithm
with $\delta_c=200$ (200c hereafter) and $\delta_c=500$ (500c
hereafter) to make a comparison between the relations calibrated with
halo samples identified by means of different methods.

\begin{figure}
\centering
    \includegraphics[width=\columnwidth]{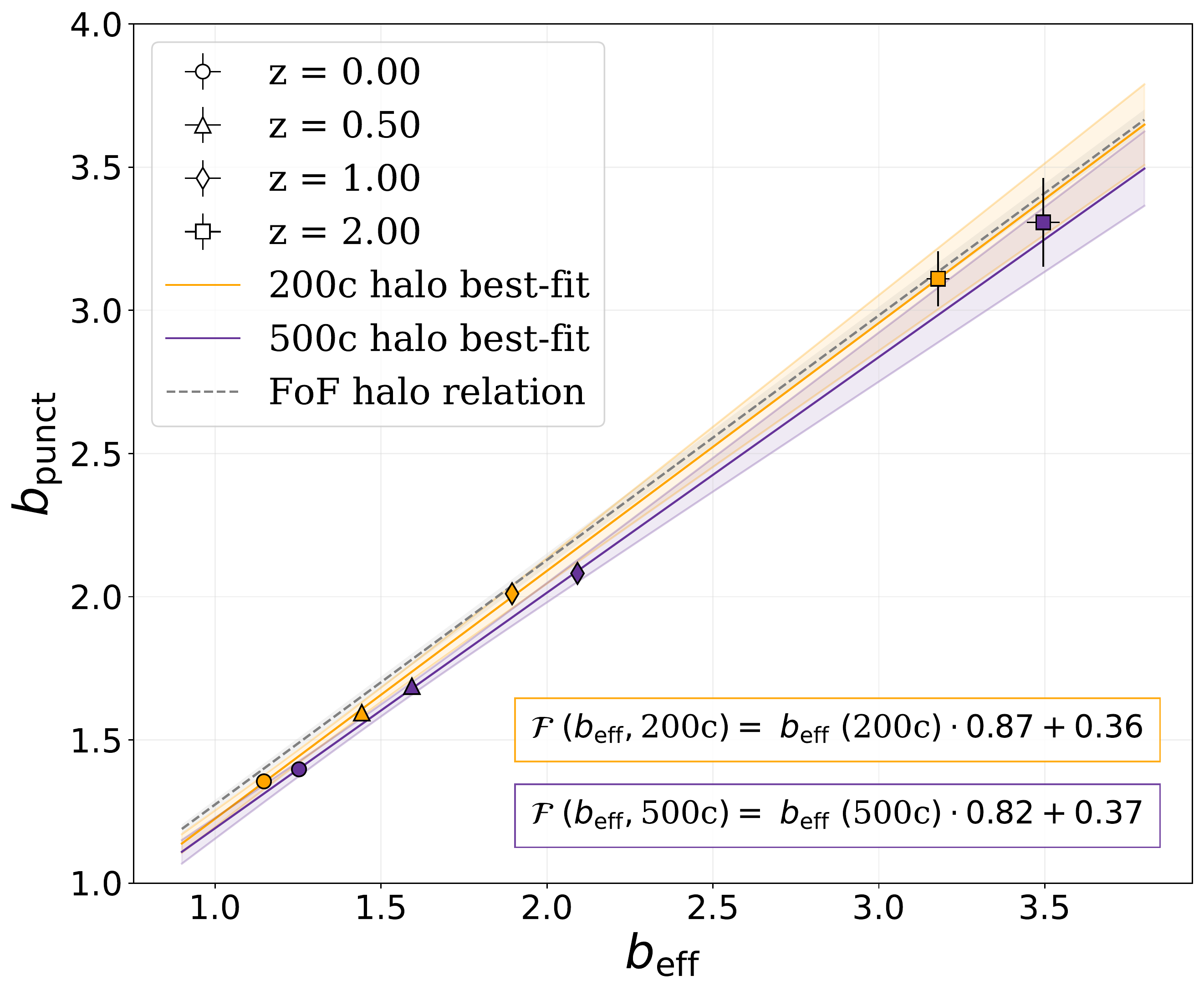}
    \caption{Linear relations between $b_\mathrm{eff}$ and
      $b_\mathrm{punct}$, calibrated using different halo catalogues
      for the standard $\Lambda$CDM scenario. The
      different markers show the data obtained in this work using 200c
      (in orange) and 500c (in violet) haloes at $z = 0, 0.5, 1,
      2$. The fitted relations are shown by solid lines and the
      corresponding relations are represented in orange and violet for
      200c and 500c, respectively. The dashed grey line represents the
      linear function calibrated in a previous work \citep{contarini2019},
      using FoF haloes, extracted from the {\small CoDECS} simulations.
      The uncertainties related to the best-fit models are reported as shaded
      regions around each linear relation.}
    \label{fig:bias_relations} 
\end{figure}
\begin{figure}
\centering
    \includegraphics[width=\columnwidth]{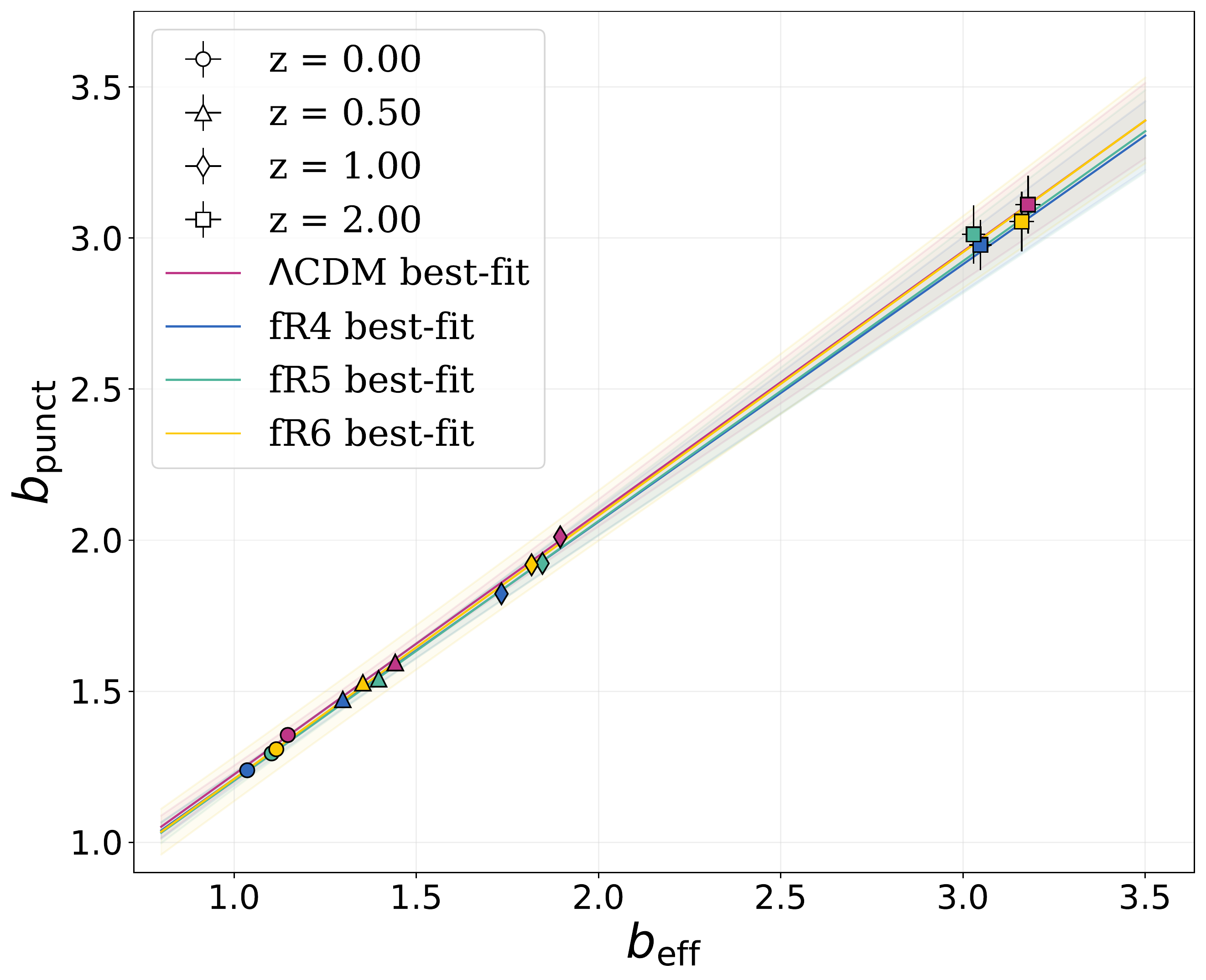}
    \caption{Linear relations between $b_\mathrm{eff}$
        and $b_\mathrm{punct}$, calibrated by means of 200c halo
        catalogues in different cosmological models. The different
        markers show the data obtained using the values computed for
        the $\Lambda$CDM (in magenta), fR4 (in blue), fR5 (in green)
        and fR6 (in yellow) models at $z = 0, 0.5, 1, 2$. The best-fit
        linear models are represented by solid lines with the same
        colours of the correspondent markers, while their
        uncertainties are reported as shaded areas around these
        lines.}
    \label{fig:bias_relations_fR} 
\end{figure}

Figure \ref{fig:bias_relations} shows the results of this
analysis. The new linear relations obtained by fitting the value of
$b_\mathrm{punct}$ as a function of $b_\mathrm{eff}$ at $z=0, 0.5, 1,
2$ are represented with different colours for the 200c and 500c
haloes. The fit obtained in \cite{contarini2019} is also displayed as
reference. We note that
going from FoF to 200c and 500c haloes, the objects we are considering
become more compact and denser. This results in a departure from the
bisector of the plane $b_\mathrm{eff}$-$b_\mathrm{punct}$,
representing the relation for matter tracers with an identical
behaviour of the bias factor on all the regions of the density
field. 
We find the following results from the fitting of the
data at different redshifts:
\begin{equation}
\label{parameters_relation}
    \mathcal{F}(b_\mathrm{eff}) = (0.87 \pm 0.02) \, b_\mathrm{eff} + (0.36 \pm 0.03) \, , 
  \text{ for 200c}
\end{equation}
and
\begin{equation}
    \mathcal{F}(b_\mathrm{eff}) = (0.82 \pm 0.02) \, b_\mathrm{eff} + (0.37 \pm 0.02) \, , 
  \text{ for 500c .}
\end{equation}
The linear function $\mathcal{F}$ shows a lowering of the slope related to the
increase of the central density selection. We can conclude that the
relation required to convert the large-scale effective bias has a
slight dependence on the selection criteria applied to define the mass
tracers, and that it has therefore to be calibrated according to the
type of objects used to identify the voids.

\begin{figure}
\centering
    \includegraphics[width=\columnwidth]{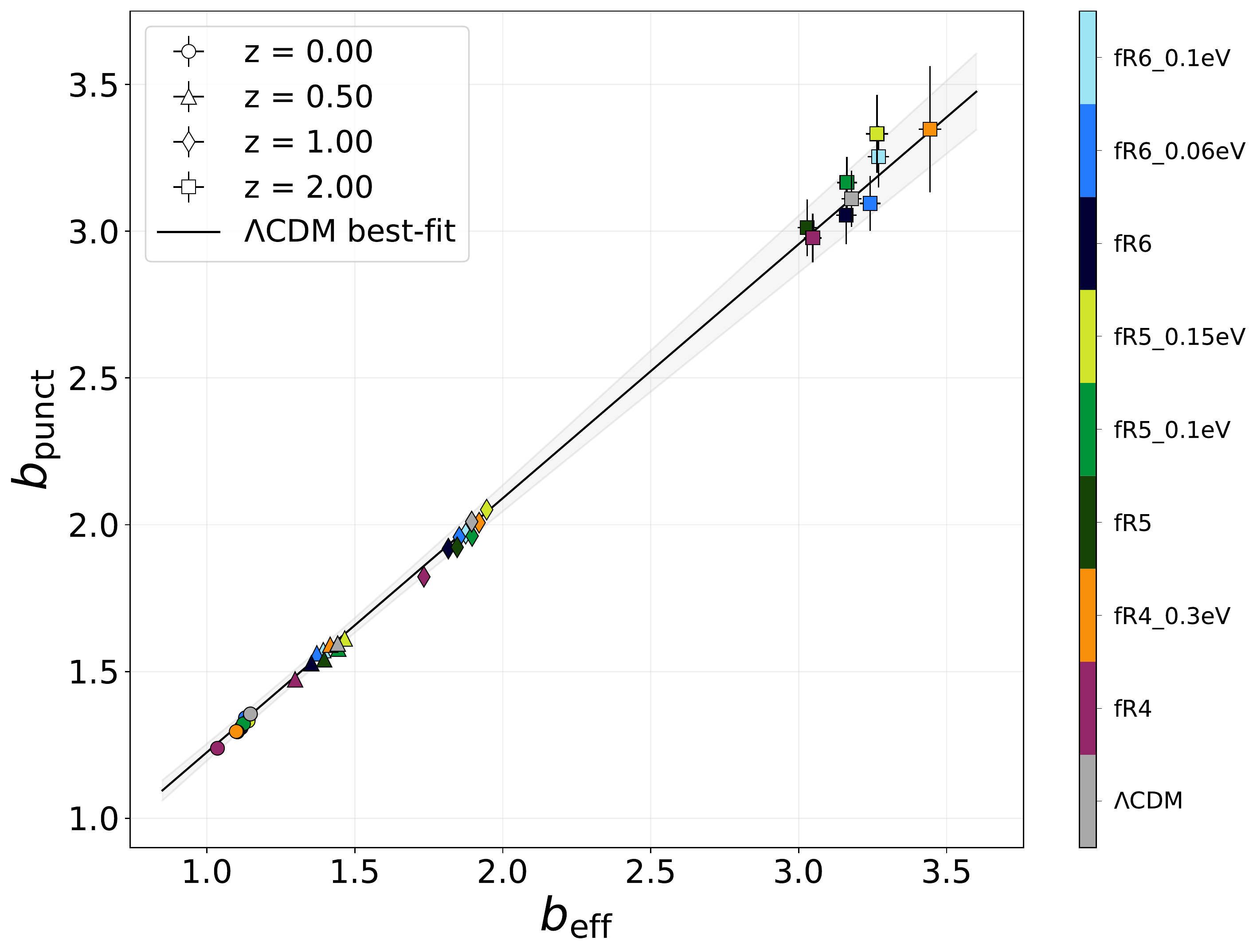}
    \caption{Values of
          $b_\mathrm{eff}$ and $b_\mathrm{punct}$ measured using 200c
          haloes at ${z = 0, 0.5, 1, 2}$, for all the cosmological
          models analysed.  The colorbar on the right reports the
          colours associated to each cosmological model. The black
          line indicates the linear relation obtained by fitting the
          $\Lambda$CDM data only, while the shaded grey region shows
          its associated uncertainty.}
    \label{fig:bias_relations_fR_ALL} 
\end{figure}

We test also the possible dependence of the function
$\mathcal{F}$ on the cosmological model. In Figure
\ref{fig:bias_relations_fR} we report the linear relations found
using the 200c haloes to compute both the values of $b_\mathrm{eff}$
and $b_\mathrm{punct}$ of the $\Lambda$CDM, fR4, fR5 and fR6
models. In this case, considering tracers with the same mass
selection, the $\mathcal{F}$ relation obtained for the $\Lambda$CDM
case results statistically indistinguishable from the ones computed
for \textit{non-standard} cosmological scenarios. We finally test
the universality of the $\mathcal{F}$ relation analysing also the
models with massive neutrinos, comparing the values of
$b_\mathrm{eff}$ and $b_\mathrm{punct}$ measured for these scenarios
using 200c haloes with those previously shown. As displayed in
Figure \ref{fig:bias_relations_fR_ALL}, the linear relation
calibrated with the $\Lambda$CDM model is fully consistent with the
data obtained for all analysed cosmological scenarios. For this
reason, in the following analysis we will apply the calibration
obtained for the $\Lambda$CDM standard scenario to obtain the
theoretical void size function for every cosmological model, that is
assuming that the $\mathcal{F}$ relation is universal, for a
specific type of tracers. In the last part of this Section we make
use of the 200c halo catalogues only, since the higher number of
tracers and the lower bias factor ease the identification of
voids. However, we tested the validity of the following methods
considering also the 500c haloes as tracers, finding consistent,
though less precise, results.

After obtaining the linear function $\mathcal{F}$ from the analysis of
both the DM particle and 200c halo density distribution, we can now
use the coefficients shown in Eq. \eqref{parameters_relation} to
convert the threshold $\delta_{v,\mathrm{tr}}^\mathrm{NL}=-0.7$.  This
density contrast is used during the cleaning procedure of voids
identified in the DM halo field and has to be properly converted to
take into account the effect of the bias factor on the theoretical
void size function. To this purpose, we first apply
Eq. \eqref{eq:thr_conversion} to obtain the nonlinear density contrast
in the DM distribution. Then we evaluate its corresponding value in
linear theory by means of Eq. \eqref{eq:bernardeau}, inserting this
quantity in the theoretical expression of the Vdn model. We repeat
this method to compute the theoretical void size function for each
cosmological scenario, using {\small MGCAMB} to obtain the matter
power spectrum, required to evaluate both the tracer effective bias,
$b_\mathrm{eff}$, and the square root of the mass variance, $\sigma
(z)$. To minimise numerical incompletenesses in the void sample, we
discard the regions with $R_\mathrm{eff}$ less than $[2.75, 2.5, 2.5,
  2.25] \times \mathrm{MPS}$ of the $\Lambda$CDM halo catalogues for
the redshifts $[0, 0.5, 1, 2]$, respectively.  In this case we did not
apply a fixed cut at small radii to reject the voids affected by
sparsity of the tracers.  Indeed, contrary to what happens with the DM
particles, the MPS of the DM haloes depends on the redshift and the
interplay between the spatial resolution of the tracers and the
incompleteness of the void number counts is not trivial.  Therefore we
prefer to apply these conservative selections relying on the drop
observed at small radii in the measured void size function at
different redshifts. This selection is a fundamental
requirement when exploiting the void size function for deriving
cosmological constraints. Indeed, it is crucial to avoid
contamination from poorly sampled spatial scales to obtain unbiased
results but, at the same time, it is important to preserve the void
number counts to avoid loss of signal and maximise the constraining
power of this statistics. We tested different minimum radius cuts
and we verified that lower values would lead to a discrepancy
between the measured and the predicted void counts, while higher
values would cause a dramatic reduction of the void statistics.
\begin{figure}
\centering
    \includegraphics[width=\columnwidth]{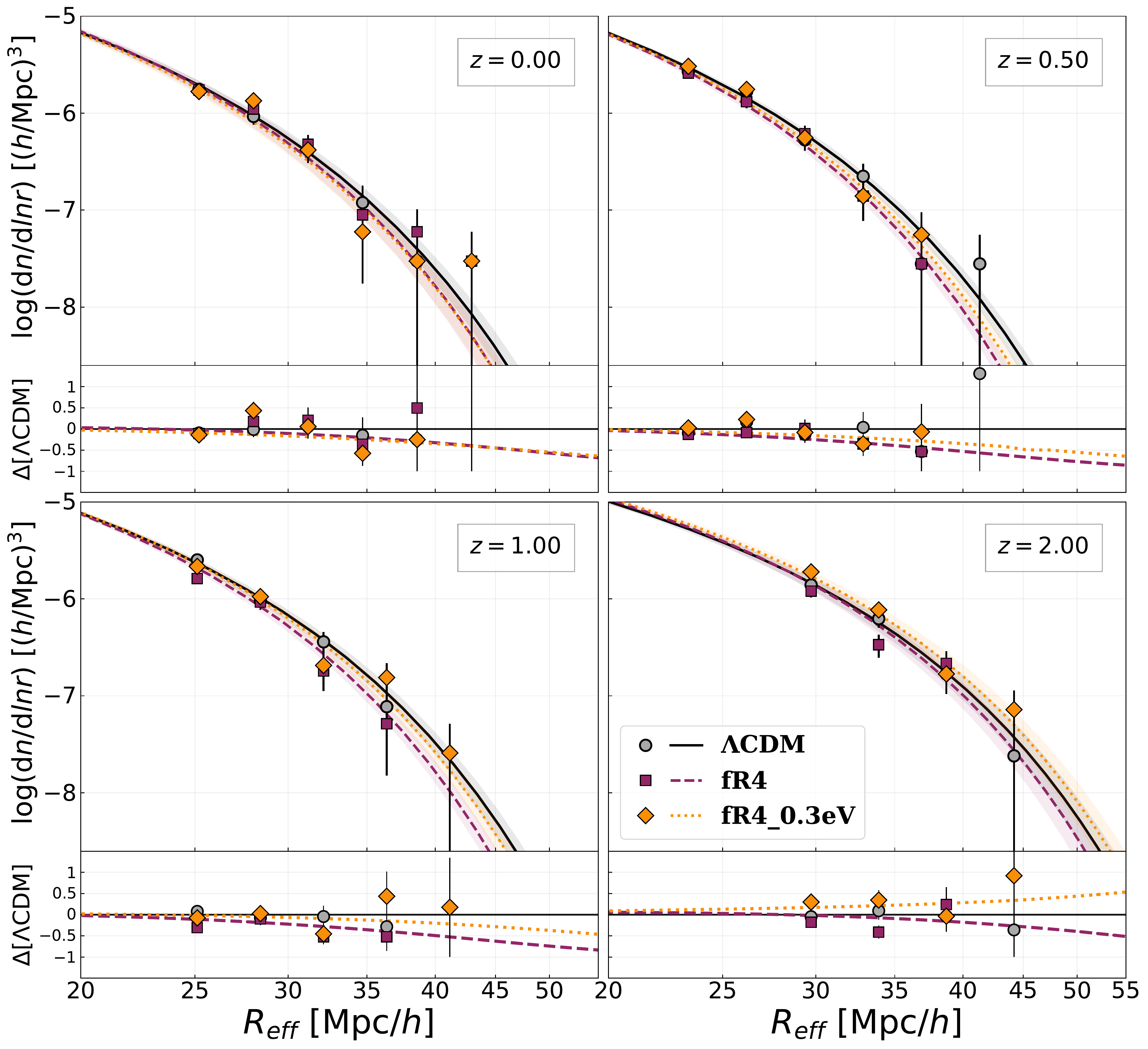}
    \caption{Measured and predicted abundances of
      cosmic voids identified in the distribution of the 200c haloes,
      extracted from the $\Lambda$CDM, fR4 and fR4\_0.3eV simulations,
      at redshifts $z=0, 0.5, 1, 2$. We represent with different
      colours and markers the abundances computed for each
      cosmological scenario, with Poissonian
      errorbars. The theoretical predictions computed
        for the considered models are reported with lines of the
        corresponding colours.  The shaded region around each curve
        represents the uncertainty given by the propagation of the
        errors during the rescaling of the Vdn underdensity threshold
        by means of the function $\mathcal{F}(b_\mathrm{eff})$.
      The bottom sub-panels report the residuals computed as the
      difference from the $\Lambda$CDM theoretical predictions,
      divided by the value of the latter, for both the
        measured and the predicted void abundances.}
    \label{fig:size_functions_haloes_fR4} 
\end{figure}
\begin{figure*}
\centering
    \includegraphics[width=1.02\columnwidth]{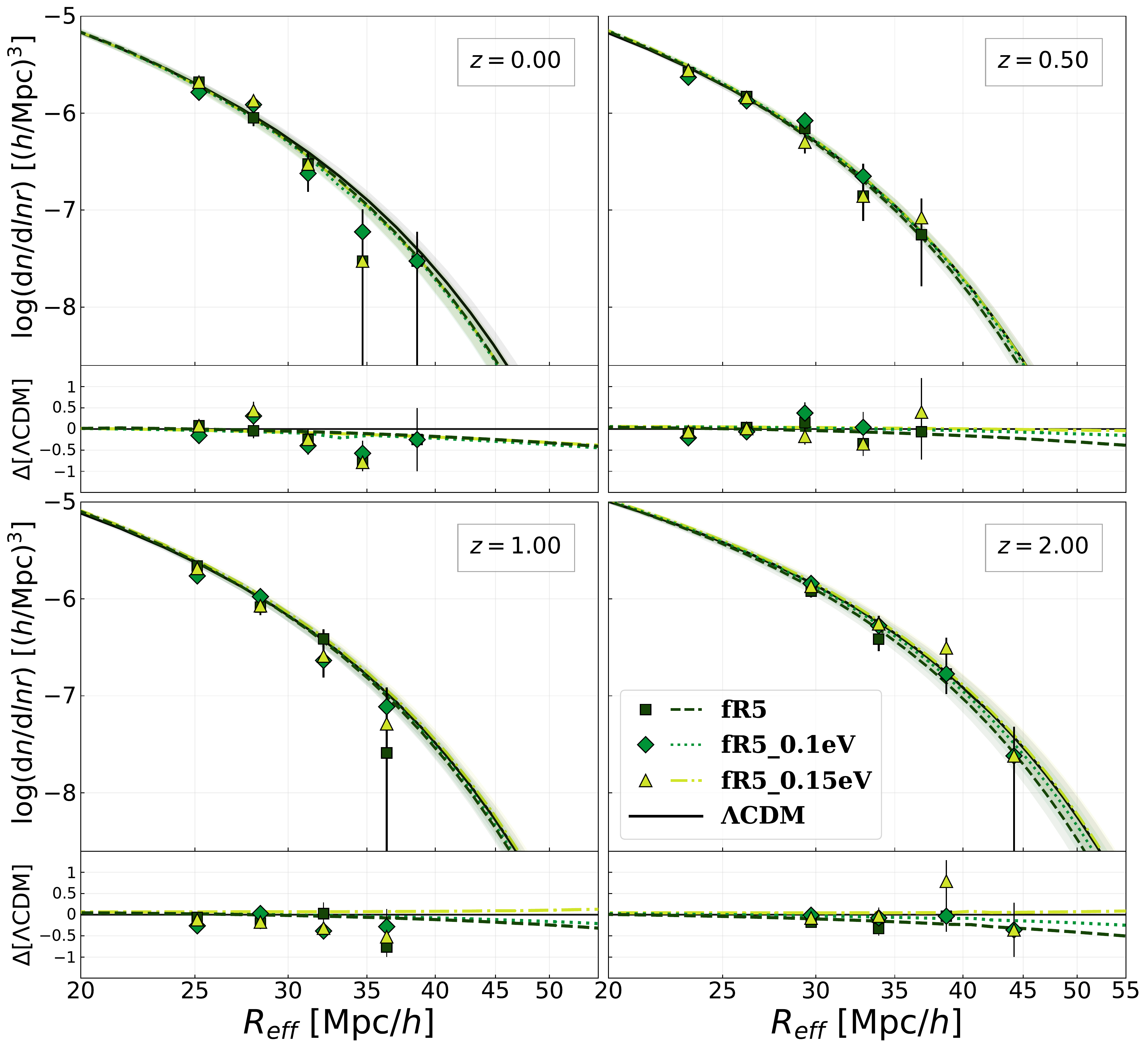}
    \includegraphics[width=1.02\columnwidth]{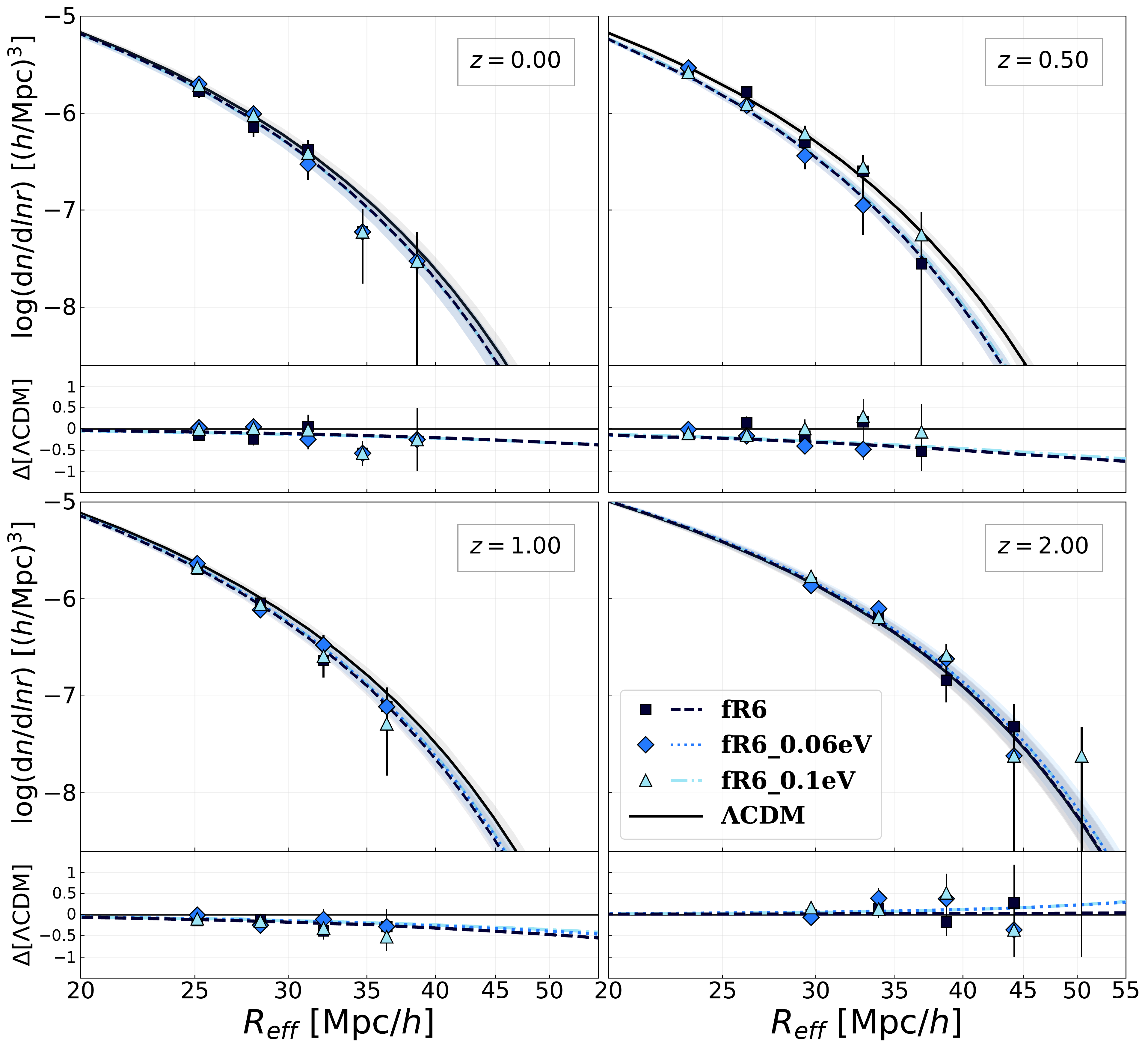}
    \caption{Measured and predicted abundances of
      cosmic voids identified in the distribution of 200c haloes,
      extracted from the fR5, fR5\_0.1eV and fR5\_0.15eV simulations
      (\textit{left panels}) and from the fR6, fR6\_0.06eV and
      fR6\_0.1eV simulations (\textit{right panels}),
        compared to the theoretical void size function for the
        $\Lambda$CDM model. In the bottom sub-panels we report the
      residuals with respect to the latter. The
        symbols are analogous to the ones reported in Fig.
        \ref{fig:size_functions_haloes_fR4}.}
    \label{fig:size_functions_haloes_fR5-6} 
\end{figure*}
In Fig. \ref{fig:size_functions_haloes_fR4} we report the comparison
between the measured void abundance for the $\Lambda$CDM, fR4 and
fR4\_0.3eV models, showing also the corresponding predictions of the
Vdn model computed for each cosmological scenario.  The shaded region
around each curve represents the uncertainty derived from the
propagation of the error associated to the value of $b_\mathrm{eff}$
computed for each case, converted by means of
Eq. \eqref{parameters_relation}, and used to compute the theoretical
models. The residuals reported in the bottom sub-panels are computed
as the difference from the theoretical void size function of the
$\Lambda$CDM model, in units of the latter, for both the measured and
the predicted abundances.  We find a good agreement between the
predictions of the reparametrised Vdn model and the measured void size
functions. Nevertheless, the cosmic voids found in the DM halo
simulations are so rare that the Poissonian noise does not allow us to
distinguish a specific trend for the abundances measured in MG and
massive neutrinos scenarios. This was previously
verified also in other works analysing voids identified in biased
tracers using cosmological simulations in MG gravity scenarios or
with massive neutrinos \citep{Voivodic2017, Kreisch2019}. In Fig.
\ref{fig:size_functions_haloes_fR5-6} we show the results for the
remaining cosmological models. Even more in these cases, the void
abundances derived in different cosmologies are hardly discernable
from the $\Lambda$CDM ones. The signal is stronger at
higher redshifts due to the fact that the underdensity threshold
used to rescale the voids moves towards values closer to $0$ for
higher values of $b_\mathrm{eff}$ (see
Eq. \eqref{eq:thr_conversion}). This causes the growth of the
population of large voids and can lead to an overall increase of the
number of voids with radii belonging to the range considered in this
analysis. However, since this method implies the selection of
shallower voids, it will be important to verify the purity of the
void sample when it derives from real galaxy surveys, and thus to
take into account the possible contamination by Poissonian
noise. The accuracy of the measured void counts at high redshifts
can be in fact compromised by systematic uncertainties still not
parameterised in the model. Nevertheless, in our case the
prescriptions adopted to prepare the void samples are proven to be
compliant with the theoretical predictions and do not require
further procedures of removal of spurious voids.

Given the poor statistics of the measured void counts
and thus the large uncertainties associated to the data, we focus
now on the analysis of the predicted abundances only. We first
point out that, although from these plots the void counts could appear
reduced in MG cosmologies, this is true in fact only for the large
sizes.  Looking at the Figures \ref{fig:size_functions_haloes_fR4} and
\ref{fig:size_functions_haloes_fR5-6}, we note that the void size
functions in the different cosmological models considered are
significantly different only for large radii, an effect that is larger
for higher values of the $f_{R0}$ parameter. While at $z=0$ the
predictions of the Vdn model for MG cosmologies with and without
massive neutrinos are statistically indistinguishable, at intermediate
redshifts the presence of massive neutrinos causes a shift of the void
size function towards the one obtained for the standard $\Lambda$CDM
model. This trend results even more relevant at $z=2$, where the
effect of the neutrino thermal free-streaming brings the theoretical
curve above the one of the $\Lambda$CDM case. This
outcome might seem counterintuitive, since the presence of massive
neutrinos leads effectively to a slow down of the evolution cosmic
voids. Nevertheless, this trend has been identified also in
\citet{Kreisch2019} using a methodology similar to the one reported
in this paper to select and characterise the void sample. This
phenomenon is in fact due to the effect of the adopted
bias-dependent threshold, that causes a rescaling of the detected
underdensities toward greater radii and appears more evident for
larger voids. This is an obvious indicator of the possibility to
use cosmic void abundances to disentangle the degeneracies between MG
and massive neutrinos models. However, we recall that
the effect of the tracer bias on the void size function may be
partially compensated by the one of massive neutrinos and MG models,
therefore the trends found in this analysis may be different using
other simulations or different tracers \citep[see
also][]{Kreisch2019}.

\begin{table*}
\centering
\caption{The most relevant quantities related to the voids identified
  in the distribution of 200c haloes. The table is structured in $4$
  parts, which separate the values computed for different
  redshifts. Each line shows the set of data relative to each of the
  cosmological models analysed in this work. The values of the tracer
  effective bias used to reparametrise the Vdn model are reported in the
  column named $b_\mathrm{eff}$. The following column provides the void
  abundances predicted by the theory, obtained by integrating the Vdn
  model over the range of void radii described in this section. Then
  we report the number of voids extracted from the void catalogues
  built with {\small VIDE}, for completeness. The last columns represent
  the void counts derived from the catalogues of voids, modelled with the
  cleaning algorithm, measured on the same range of void radii used to
  compute the theoretical abundances.}
\begin{tabular}{lcccc|cccc}
\hline
\toprule
Cosmological model & $b_\mathrm{eff}$ & Vdn model & {\small VIDE} voids  
& Cleaned voids & $b_\mathrm{eff}$  & Vdn model & {\small VIDE} voids  & Cleaned voids \\
\midrule
 & 
\multicolumn{4}{c}{$z=0$} & \multicolumn{4}{c}{$z=0.5$} \\
\noalign{\smallskip}
\cline{2-9}
\noalign{\smallskip}
$\Lambda$CDM &  $1.147 \pm 0.009$ & $119 \pm 16$ & $703 \pm 27$ & $109 \pm 10$ &
$1.44 \pm 0.01$ & $193 \pm 17$ & $949 \pm 31$ & $185 \pm 14$ \\  
fR4 &  $1.036 \pm 0.008$ & $111 \pm 19$ & $682 \pm 26$ & $117 \pm 11$ & 
$1.30 \pm 0.01$ & $163 \pm 18$ & $912 \pm 30$ & $168 \pm 13$ \\ 
fR4\_0.3eV & $1.099 \pm 0.009$ & $104 \pm 16$ & $703 \pm 27$ & $119 \pm 11$ & 
$1.42 \pm 0.01$ & $173 \pm 17$ & $941 \pm 31$ & $199 \pm 14$ \\ 
fR5 & $1.103 \pm 0.009$ & $116 \pm 17$ & $690 \pm 26$ & $112 \pm 11$ & 
$1.40 \pm 0.01$ & $192 \pm 17$ & $919 \pm 30$ & $179 \pm 13$ \\
fR5\_0.1eV & $1.124 \pm 0.009$ & $112 \pm 16$ & $679 \pm 26$ & $107 \pm 10$ & 
$1.44 \pm 0.01$ & $199 \pm 17$ & $930 \pm 30$ & $170 \pm 13$ \\ 
fR5\_0.15eV & $1.140 \pm 0.009$ & $113 \pm 16$ & $683 \pm 26$ & $126 \pm 11$ & 
$1.47 \pm 0.01$ & $201 \pm 17$ & $922 \pm 30$ & $177 \pm 13$ \\
fR6 & $1.116 \pm 0.008$ & $108 \pm 16$ & $702 \pm 26$ & $96 \pm 8$ & 
$1.35 \pm 0.01$ & $149 \pm 15$ & $907 \pm 30$ & $180 \pm 13$ \\
fR6\_0.06eV & $1.131 \pm 0.009$ & $107 \pm 15$ & $715 \pm 27$ & $113 \pm 11$ & 
$1.37 \pm 0.01$ & $148 \pm 15$ & $900 \pm 30$ & $165 \pm 13$ \\
fR6\_0.1eV & $1.140 \pm 0.009$ & $106 \pm 15$ & $723 \pm 27$ & $113 \pm 11$ & 
$1.39 \pm 0.01$ & $153 \pm 15$ & $929 \pm 30$ & $173 \pm 13$ \\ 
\hline
\noalign{\smallskip}
 & 
\multicolumn{4}{c}{$z=1$} & \multicolumn{4}{c}{$z=2$} \\
\noalign{\smallskip}
\cline{2-9}
\noalign{\smallskip}
$\Lambda$CDM & $1.90 \pm 0.01$ & $154 \pm 17$ & $734 \pm 27$ & $152 \pm 12$ & 
$3.18 \pm 0.03$ & $97 \pm 17$ & $470 \pm 22$ & $92 \pm 10$ \\ 
fR4 & $1.73 \pm 0.01$ & $129 \pm 13$ & $727 \pm 27$ & $108 \pm 10$ & 
$3.05 \pm 0.03$ & $92 \pm 15$ & $435 \pm 21$ & $73 \pm 9$ \\ 
fR4\_0.3eV & $1.92 \pm 0.02$ & $147 \pm 15$ & $731 \pm 27$ & $140 \pm 12$ & 
$3.44 \pm 0.04$ & $114 \pm 18$ & $469 \pm 22$ & $121 \pm 11$ \\
fR5 & $1.85 \pm 0.01$ & $154 \pm 16$ & $750 \pm 27$ & $133 \pm 12$ & 
$3.03 \pm 0.03$ & $85 \pm 14$ & $460 \pm 21$ & $73 \pm 9$ \\
fR5\_0.1eV & $1.90 \pm 0.01$ & $155 \pm 16$ & $743 \pm 27$ & $120 \pm 11$ & 
$3.16 \pm 0.03$ & $93 \pm 16$ & $477 \pm 22$ & $90 \pm 9$ \\
fR5\_0.15eV & $1.95 \pm 0.02$ & $164 \pm 17$ & $742 \pm 27$ & $125 \pm 11$ & 
$3.26 \pm 0.04$ & $101 \pm 17$ & $479 \pm 22$ & $93 \pm 10$ \\
fR6 & $1.82 \pm 0.01$ & $132 \pm 15$ & $744 \pm 27$ & $125 \pm 11$ & 
$3.16 \pm 0.03$ & $98 \pm 17$ & $456 \pm 21$ & $96 \pm 10$ \\
fR6\_0.06eV & $1.85 \pm 0.01$ & $136 \pm 16$ & $737 \pm 27$ & $136 \pm 12$ & 
$3.24 \pm 0.04$ & $103 \pm 17$ & $484 \pm 22$ & $101 \pm 10$ \\
fR6\_0.1eV & $1.87 \pm 0.01$ & $136 \pm 16$ & $742 \pm 27$ & $128 \pm 11$ & 
$3.27 \pm 0.04$ & $103 \pm 17$ & $468 \pm 22$ & $111 \pm 11$ \\
\hline
\bottomrule
\end{tabular}
\label{tab:total_abundance_comparison}
\end{table*}

To facilitate the comparison of these results and maximise the signal
obtained from the measured void abundance, we compare now the total
void number counts with the abundance computed by integrating the
theoretical void size function over the same range of
radiiEven if the total void counts is not commonly
used to derive cosmological constraints, it can constitute in this
case a useful quantity to analyse. Indeed, it allows to perform a
simple validation of the predictions of the void size function
models, collapsing the information on void number counts related to
different spatial scales and sharping the signal achieved from the
measured abundances. We present in Table
\ref{tab:total_abundance_comparison} the comparison between
the integrated values of void number counts obtained
from the theoretical models and our measurements, for each of the
cosmological models and redshifts explored in this work. We report
also the value of the tracer effective bias, used to reparametrise the
characteristic threshold of the Vdn model. Then we show, for
completeness, the measured abundance derived from the {\small VIDE}
void catalogues before performing the cleaning
procedure\footnote{Since the {\small VIDE} void radii are
systematically larger than the ones rescaled by means of the
cleaning algorithm described in Section
\ref{sec:finding_and_cleaning}, we applied a more severe cut to
discard the voids affected by the sparsity of the tracers. In
particular, to minimise the numerical incompleteness for small
radii, we increase the minimum radius of the accepted voids by a
factor of $1.5$ with respect to the selection adopted for the
cleaned catalogues.}. The abundances extracted from the raw {\small
VIDE} void catalogues are significantly larger than those obtained
after the cleaning procedure, but they are clearly not in agreement
with the Vdn model predictions. This outcome is not surprising since
these voids are not modelled according to the excursion-set theory
described in Section \ref{sec:void_size_function_theory}.  Now we
focus on the comparison of the total counts of cleaned voids with the
predictions achieved with the reparametrised Vdn model. The errors
associated with the latter are evaluated by propagating the
uncertainties related to $b_\mathrm{eff}$ and $b_\mathrm{punct}$
during the calibration of the function $\mathcal{F}(b_\mathrm{eff})$,
while those associated with the measured void abundances are assumed
to be Poissonian. We can see that the theoretical void abundances are
overall consistent with the observed ones, considering the
uncertainties on both the values. As expected from the results shown
in Figures \ref{fig:size_functions_haloes_fR4} and
\ref{fig:size_functions_haloes_fR5-6}, a significant
differentiation between the analysed cosmological models is reached at
$z=2$, despite the scarcity of void counts makes their distinction
challenging. Indeed, the simulations considered in this work do not
allow us to have enough statistics for large voids at high redshifts.


\section{CONCLUSIONS}
\label{sec:conclusions}

In this work we have investigated the possibility of disentangling the
degeneracies characterising cosmological models that simultaneously
feature a modification of GR -- in the form of $f(R)$ gravity -- and
the presence of massive neutrinos. To explore possible observational
differences among these scenarios, we have focused on the exploitation
of cosmic void density profiles and abundances. We have built void
catalogues by means of the void detection algorithm {\small VIDE},
identifying voids in both the DM particle and halo distributions, for
all the different cosmological models at our disposal, at the
redshifts $z=0, 0.5, 1, 2$. For the analysis of the void size
function, we adopted the procedure described in \citet{contarini2019},
modelling cosmic voids according to the theoretical definition
provided by the Vdn model.

The main results obtained in this work can be summarised as follows:
\begin{itemize}
    \item The analysis of the void stacked density profiles, measured in both the DM and halo distributions, has revealed some deviations of the MG model profiles from those computed in a $\Lambda$CDM cosmology, more evident for the most extreme scenarios. These differences are particularly strong at $z=1$, when the growth of cosmic structures shows an enhancement given by the effect of the fifth force. Nevertheless, the neutrino thermal free-streaming almost completely erases any peculiar trend of the density profiles, making void profiles measured for these models almost indistinguishable compared to the $\Lambda$CDM ones.
    \item We have found an excellent agreement between the measured
        abundances and the theoretical predictions obtained with
        the Vdn model for voids identified in the DM
        particle distribution in different cosmological
        models. Furthermore, while at low redshifts the presence
        of massive neutrinos tends to lower the void 
        size function computed for MG models towards the one relative
        to the $\Lambda$CDM scenario, at high redshifts this effect results in an excessive reduction in the void abundance.
        This inversion of the trend is caused by the different redshift dependence of MG and massive neutrinos imprints on structure formation. In fact, the effect of massive neutrinos to damp the
        evolution of voids is already in place at early epochs when MG effects are still negligible.
    \item We have fitted a linear function $\mathcal{F}$ to model the
        relation between the tracer bias computed on large scales,
        $b_\mathrm{eff}$, and the one measured inside cosmic voids,
        $b_\mathrm{punct}$. We have considered two
        selection criteria to build samples of viralised DM haloes
        with different compactness, characterised by an internal
        density equal to $200$ and $500$ times the critical density of
        the Universe, respectively. From the comparison between the
        coefficients of the function $\mathcal{F}$ obtained using
        these different types of biased tracers, we have identified
        the presence of a trend characterising the relation between
        $b_\mathrm{eff}$ and $b_\mathrm{punct}$. This appears as a
        slight dependence on the type of objects used to identify voids,
        related in particular to the criterion applied
        to define the mass tracers. We have also tested
        the universality of the calibrated relation  for a specific
        selection of mass tracers. In particular, using 200c haloes,
        we have compared the linear function obtained for $\Lambda$CDM
        model with those computed using \textit{non-standard}
        cosmological models, verifying that the calibrations performed
        for the different scenarios are statistically consistent.
    \item With the parametrisation of the threshold of the Vdn model
        by means of the function $\mathcal{F}(b_\mathrm{eff})$, we have compared
        the measured and predicted abundances of voids identified in the
        200c halo catalogues. We have found a good agreement
        between the void size functions measured in the simulated void
        catalogues and the predicted ones, in the full range of void
        radii probed by our simulations. For these sizes, all the
        cosmological models considered in this work predict
        statistically indistinguishable void abundances.  Larger
        simulations are required to push the analysis at larger voids,
        where the differences in the size function of cosmic
        voids are expected to be larger, thus allowing
        to break the cosmic degeneracies.
\end{itemize}
We can conclude that the void density profiles do not allow to
disentangle the cosmic degeneracies given by the proper combination of
the $f_{R0}$ and $m_v$ parameters. On the other hand, void abundances
have been shown to be a promising probe to break these degeneracies,
though larger simulation volumes are needed to extract more precise
and accurate results for bigger voids at higher redshifts. The
requirement for a proper exploitation of this probe is therefore the
exploration of wide and deep regions of the Universe, with the goal of
obtaining a statistically relevant number of voids with large radii,
at $z>1$. Future spectroscopic surveys like WFIRST, \textit{Euclid}
and LSST, will serve to this purpose, allowing to achieve the cosmic
void statistics required to disentangle the degenerate effects of MG
and massive neutrino scenarios.

\section*{Data Availability}
The simulation data underlying this article will be shared on reasonable request to the corresponding author.

\section*{Acknowledgements}
SC is particularly grateful to Alice Pisani for her useful comments and suggestions.
FM, LM, CG and MB acknowledge the grants ASI n.I/023/12/0, ASI-INAF n. 2018-23-HH.0 and PRIN MIUR 2015 ``Cosmology and Fundamental Physics: illuminating the Dark Universe with Euclid''. 
LM and CG acknowledge support from PRIN MIUR 2017 WSCC32 ``Zooming into dark matter and proto-galaxies with massive lensing clusters''.
MB also acknowledges support by the project "Combining Cosmic Microwave Background and Large Scale Structure data: an Integrated Approach for Addressing Fundamental Questions in Cosmology", funded by the MIUR Progetti di Ricerca di Rilevante Interesse Nazionale (PRIN) Bando 2017 - grant 2017YJYZAH. The N-body simulations described in this work have been performed on the Hydra supercomputer at RZG and on the Marconi supercomputer at Cineca thanks to the PRACE allocation 2016153604 (P.I. M. Baldi). The authors acknowledge the use of computational resources
from the parallel computing cluster of the Open Physics
Hub (\url{https://site.unibo.it/openphysicshub/en}) at the
Physics and Astronomy Department in Bologna.



\bibliographystyle{mnras}
\bibliography{main}



\bsp	
\label{lastpage}
\end{document}